\documentclass[aps,10pt,prx,twocolumn,showpacs,amsmath,amssymb,superscriptaddress]{revtex4-1}

\usepackage{graphicx}
\usepackage{hyperref}
\usepackage{txfonts}
\usepackage{amsmath}
\usepackage{xcolor}

\newcommand{\lh}{{\text{lh}}}
\newcommand{\hh}{{\text{hh}}}
\newcommand{\cb}{{\text{cb}}}

\newcommand{\alphabar}{{\bar{\alpha}}}

\newcommand{ \bra }			[1] { \left<{#1}\right| }
\newcommand{ \ket }				[1] { \left|{#1}\right\rangle }
\newcommand{ \kets }				[1] { |{#1}\rangle }
\newcommand{ \braket }				[1] { \left\langle{#1}\right\rangle }
\newcommand{ \brakets }				[1] { \langle{#1}\rangle }

\begin{document}

\title{Detecting nonlocal Cooper pair entanglement by optical Bell inequality violation}

\author{Simon E. Nigg}
\affiliation{Department of Physics, University of Basel, Klingelbergstrasse 82, 4056 Basel, Switzerland}
\author{Rakesh P. Tiwari}
\affiliation{Department of Physics, University of Basel, Klingelbergstrasse 82, 4056 Basel, Switzerland}
\author{Stefan Walter}
\affiliation{Department of Physics, University of Basel, Klingelbergstrasse 82, 4056 Basel, Switzerland}
\affiliation{Institute for Theoretical Physics, University Erlangen-N{\"u}rnberg, Staudtstr. 7, 91058 Erlangen, Germany}
\author{Thomas L. Schmidt}
\affiliation{Department of Physics, University of Basel, Klingelbergstrasse 82, 4056 Basel, Switzerland}
\affiliation{Physics and Materials Science Research Unit,
University of Luxembourg, L-1511 Luxembourg}

\date{\today}

\begin{abstract}
Based on the Bardeen Cooper Schrieffer (BCS) theory of superconductivity, the coherent splitting of Cooper pairs from a
superconductor to two spatially separated quantum dots has been predicted to
generate nonlocal pairs of entangled electrons. In order to test this
hypothesis, we propose a scheme to transfer the spin state of
a split Cooper pair onto the polarization state of a pair of optical
photons. We show that the
produced photon pairs can be used to violate a Bell inequality, unambiguously demonstrating
the entanglement of the split Cooper pairs.
\end{abstract}
\maketitle

\section{Introduction}
Entanglement~\cite{Schroedinger-1935a}, i.e., correlations between parts of a quantum system
that defy any classical description, lies at the heart of quantum mechanics. It is the basis for many applications of quantum information
theory, such as quantum teleportation~\cite{Bouwmeester-1997a},
quantum computing~\cite{Jozsa-2003a}, quantum
cryptography~\cite{Jennewein-2000a}, and quantum metrology~\cite{Giovannetti-2011a}.
The first experimental demonstration of entanglement
has been achieved by violating Bell's
inequality~\cite{Bell-1964a} with polarization-entangled optical
photon pairs generated during spontaneous parametric down-conversion in a
nonlinear crystal~\cite{Aspect-1982a}. In many
applications, it is desirable to have a source of entangled pairs of spatially separated
particles. Such pairs are called EPR pairs
in reference to the seminal work of Einstein, Podolsky
and Rosen on the completeness of quantum mechanics~\cite{Einstein-1935a}.

Compared to quantum optical scenarios, the generation of \emph{electronic} EPR pairs
is rather challenging. EPR pairs of electrons are nonetheless highly desirable because an on-demand generation of such pairs would facilitate certain quantum
communication tasks in solid-state devices~\cite{Bennett2000}. Theoretically, a conventional $s$-wave superconductor provides
a natural source for electronic EPR pairs \cite{choi00,recher01,Lesovik-2001a,Hofstetter2009,Herrmann-2010a,Das2012,cottet12,Braunecker-2013a,cottet14}: the electrons in a BCS superconductor form spin singlet
Cooper pairs in the ground state.
Following theoretical proposals~\cite{recher01, Lesovik-2001a}, the coherent splitting of Cooper pairs, originating from a superconducting electrode, into two spatially
separated electrons on neighboring quantum dots (QDs) has recently
been demonstrated experimentally~\cite{Hofstetter2009, Herrmann-2010a,
Das2012}. 
While measurements of the current flowing out of the QDs have indeed demonstrated the splitting of Cooper
pairs, the detection of the spin entanglement of the expected electronic singlet state has so far remained elusive. 
Similar devices can be used as a tool to detect unconventional pairing in superconductors~\cite{Tiwari2014} or to entangle mechanical resonators~\cite{Walter2013}.

Detecting the entanglement of electronic
EPR pairs is not as straightforward as it is with their counterparts in quantum optics. Several works~\cite{Kawabata-2001a,Chtchelkatchev-2002a,Braunecker-2013a,Scherubl-2014a}
propose to violate a Bell-type inequality with current noise measurements. However, this will require accurate measurements of the
cross-correlations between the currents from the two QDs. A measurable signal only emerges if
these currents are large enough, i.e., for a strong coupling of the QDs to the measurement device
(``open quantum dots''). This conflicts with the requirement of isolating the QDs from the environment
(``closed quantum dots''), which is necessary for splitting Cooper pairs coherently in the first place.
Moreover, most existing proposals involve the use of strong ferromagnets
and complex sample geometries, and neglect (possibly long-range) electron-electron interactions when computing the current-current
correlations. As shown in Ref.~\cite{Burkard-2000a}, such interactions can reduce the
measured entanglement signal. Another approach, which is closest in
spirit to our work, is taken in~\cite{Cerletti-2005a,Budich-2010a}. These authors
investigate the possibility to transfer electron spin entanglement to
photon polarization states. These particular schemes however, suffer from a low
detection efficiency and require the use of additional quantum
resources to generate a pure two-photon state~\cite{footnote}. Very recently, the possibility of generating polarization entangled photons in a superconducting p-n junction has been proposed~\cite{Schroer2014}.

\begin{figure}[t]
\includegraphics[scale=0.9, viewport=0 0 253 140]{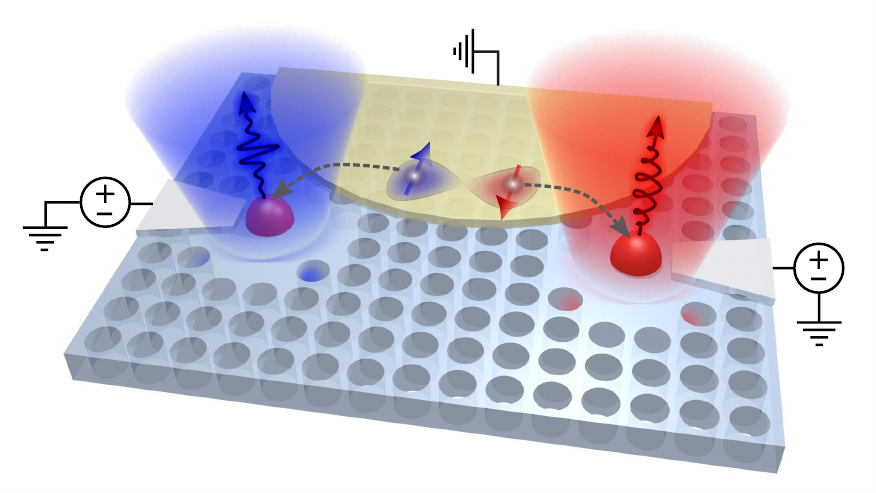}
\caption{Schematics
  of a possible realization of the entanglement transfer scheme. The
  perforated blue slab represents a photonic crystal with two
  cavities (central areas without holes). Photons in each cavity are
  coupled to the respective QDs (red domes) via electric dipole
  interactions. The QDs are tunnel coupled to a superconductor (yellow half-disc on
  top). Gates (gray slabs) allow for tuning of the QD chemical
  potentials. The emission cones of the entangled photons are depicted as red and blue shades.\label{fig:setup}}
\end{figure}

Our proposal avoids the above difficulties by converting the spin
entanglement of a {\em single} Cooper pair
into polarization entanglement of a {\em single} pair of optical
photons and requires only classical resources such as laser drives and
tunable gate voltages. Because photons do not interact with each
other, photonic entanglement is more robust to perturbations than electronic entanglement, and can be detected using standard Bell-type
measurements. Provided sufficiently independent cavities are used
(See Appendix.~\ref{sec:effect-cavity-cross} and Ref.~\cite{Brossard-2013a}), there can be no doubt that the entanglement
ultimately measured stems from the split Cooper pair because our entanglement transfer scheme involves only local
operations. Since our scheme
does not involve the measurement of electronic currents, it
works even in closed QDs. Finally, we show that the entanglement
transfer can be carried out on time scales small compared to $T_2$,
the intrinsic coherence time of the QDs (see Appendix~\ref{sec:typic-valu-param}).

\section{Setup}
Let us first present our proposed experimental setup in more
detail. A schematic drawing of a possible realization is shown in Fig.~\ref{fig:setup}. Our starting point is the typical setup for Cooper pair
splitters, i.e., a superconductor which is tunnel coupled to two
nearby QDs \cite{recher01}. The
spacing between the QDs should be smaller than the superconducting
coherence length and the QDs are assumed to be in the Coulomb
blockade regime such that adding an electron to the QDs requires a
large charging energy $U$. The onsite energies of the QDs can be
tuned via gate voltages. Splitting a Cooper pair into a singlet
state shared between the two QDs becomes energetically possible if
the total energy of the singlet state coincides with the chemical
potential of the superconductor. Both QDs are embedded into optical cavities that serve as
frequency filters allowing only certain desired
optical transitions. The small distance between the
QDs rules out conventional optical cavities, but photonic crystal
cavities are nowadays easy to manufacture at the required
length scales and can have optical linewidths and frequencies
compatible with our proposal \cite{Yoshie-2004a}. Moreover,
cavities with high quality factors and directional
out-coupling of photons into a narrow solid angle for high-efficiency
collection have been fabricated~\cite{Kim-2006a,Tran-2009a,Portalupi-2010a} and self-assembled QDs
have been successfully embedded into photonic crystals in several
experiments~\cite{Yoshie-2004a,Majumdar-2013a, Sweeney-2014a}.

\section{Entanglement transfer scheme}
We will now present our scheme for transferring the spin entanglement of a Cooper pair onto the
polarization state of a photon pair. For simplicity, we discuss a
left-right symmetric setup. To be specific, we assume that the QDs are realized as self-assembled GaAs QDs.
The relevant electronic
levels are thus generated from the light-hole ($\lh$) and heavy-hole ($\hh$) bands forming
the valence band, as well as the conduction band ($\cb$). The energy
difference between the hole bands $\Delta E = E^{\hh} - E^{\lh}$
is of the same order as the superconducting gap $\Delta$, whereas the
transition frequency between the valence band and the conduction band is in the optical frequency range \cite{Gywat-2010}. Moreover, we assume that a weak magnetic field is applied which causes a Zeeman splitting $\Delta_Z$ (with $|\Delta_Z| \ll \Delta, \Delta E$) of all electronic levels.

\begin{figure}[t]
\begin{center}
\includegraphics[viewport=0 0 253 128]{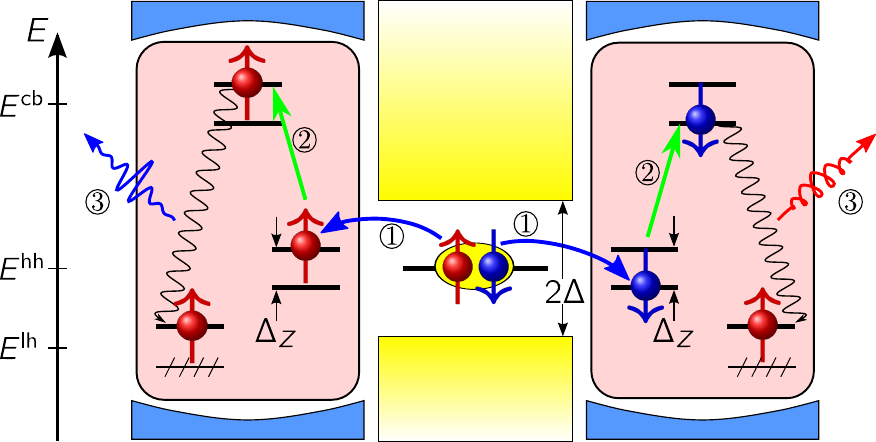}
\caption{Level diagram and schematics of the three phases of the
  entanglement transfer. The lowest light-hole states $\ket{\lh,\downarrow}_{L,R}$ are
  occupied and transitions into them are forbidden because of the Pauli principle (hash marks).
\label{fig:setupSchematics}}
\end{center}
\end{figure}

The entanglement transfer can be split into initialization and three
phases, which we discuss next. A schematic level diagram along with the essential steps of our
scheme is shown in Fig.~\ref{fig:setupSchematics}.

\emph{Initialization:}
Initially, the gate voltages of the QDs are tuned in such a way that the lowest light-hole states
on each QD, $\ket{\lh,\downarrow}_{L}$ and
$\ket{\lh,\downarrow}_{R}$, are occupied.
Furthermore the heavy-hole level resides in the superconducting gap but is detuned
with respect to the chemical potential of the superconductor.

\emph{Phase 1:}
The splitting of a Cooper pair is achieved by tuning the gate voltages
to bring the heavy-hole level into resonance with the
chemical potential of the superconductor. Single-particle tunneling is
suppressed due to the large superconducting gap. Furthermore, the large
onsite Coulomb interaction in the QD suppresses the
tunneling of both electrons of a Cooper pair onto the same QD. The
Cooper pair splitting process, where one electron tunnels to each QD,
is thus the dominant process~\cite{recher01}. When the
separation between the QDs is much smaller than the superconducting
coherence length, the Cooper pair splitting rate is (see Appendix~\ref{supp:SW})
$\hbar\Gamma_c\approx\pi\rho_0w_Lw_R[1-(\Delta_Z/2\Delta)^2]^{-1/2}$.
Here, $w_L$ ($w_R$) denotes the electronic tunnel amplitude
between the superconductor and the
left (right) QD and $\rho_0$ denotes the normal-state density of
states of the superconductor. If $\Gamma_c\gg 1/T_2$, where $T_2$ is
the intrinsic coherence time of the QDs, this
process is coherent and leads to Rabi oscillations between the superconductor
and the heavy-hole states on the QDs.
Ideally, after half a period, the double-QD is occupied by a singlet state and the
oscillation is stopped by detuning the heavy-hole level away from
resonance.

\emph{Phase 2:}
Next, the electrons in the QDs are excited from the
heavy-hole to the conduction band. This is achieved by switching
on a strong drive laser on each of the two
QDs, with frequency $\hbar \omega_{\rm drive} \approx E^{\cb} - E^{\hh}$.
A linearly polarized drive can be used, which induces
spin conserving transitions. After half a Rabi period the laser is switched off, having lifted the singlet state into the
conduction band levels. If the Zeeman splittings of the heavy hole
and conduction bands differ, one may use two drive lasers per QD to satisfy
the resonance conditions for the two different transition
frequencies simultaneously. The duration of this step is inversely
proportional to the drive strength and can thus be made fast compared with $T_2$.

\emph{Phase 3:}
The resonance frequencies of the two optical cavities are chosen to be close to the transition frequency
between the conduction band and the light-hole band, $\hbar \omega_{\rm cav} \approx E^{\cb} - E^{\lh}$.
Furthermore, the cavity linewidth $\kappa$ is assumed to be much
smaller than the frequency separation
between the light- and heavy-hole bands, i.e., $\kappa\ll (E^{\hh}-E^{\lh})/\hbar$. Therefore, the decay of the conduction
band electrons into the light-hole band due to the dipole
coupling of strength $\hbar g$ between electrons and photons, will be strongly enhanced, whereas the decay into heavy-hole states is suppressed. Since the lowest
light-hole state $\ket{\lh,\downarrow}$ is always occupied, a conduction band electron in the states
$\ket{\cb,\uparrow}$ or $\ket{\cb,\downarrow}$ can only transition to the empty $\ket{\lh,\uparrow}$
state via the emission of a linearly or circularly polarized
photon, respectively.

To investigate these three phases, we have numerically solved the
Schr\"odinger equation for the full system in the coherent limit
$\kappa\rightarrow 0$ (see Appendix~\ref{sec:numerical-simulation}). Figure~\ref{fig:CombinedPlot} shows the evolution during each phase of the occupation of the electronic levels
and the cavity mode for an optimal choice of the drive strengths and drive durations. In the limit $\kappa \to 0$, the
emitted photons
undergo coherent
oscillations between the QD and the cavity. Ideally, after half a Rabi period
$\approx \pi/(2g)$, the electrons in both QDs occupy the $\ket{\lh,\uparrow}$ states
while the electronic entanglement has been transferred to the photons.
\section{Photon extraction and Bell test}
In a real experiment, the photons need to be
extracted from the cavities for measurement. This is achieved by coupling each cavity to
a continuum of modes, e.g., as provided by a waveguide. Hence, the cavity
acquires a finite loss rate $\kappa > 0$. As discussed further
below, we will focus on the weak coupling limit, where
$g\ll \kappa$. In this limit, the coherent oscillations are suppressed
and the photons are emitted into the continuum
on a time scale $\propto \kappa/g^2$.

\begin{figure}[t]
\begin{center}
\includegraphics[viewport=1 2 250 287]{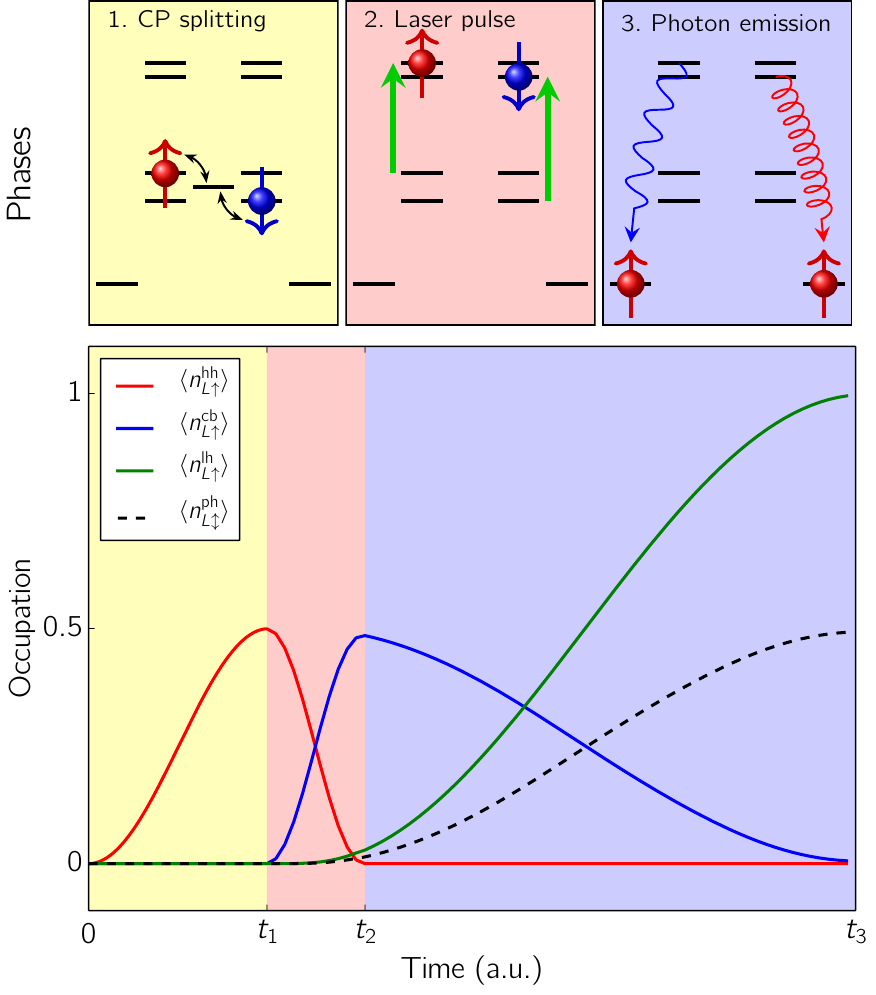}\caption{\emph{Upper panel:} Schematics of the three phases showing Cooper pair splitting in phase one, laser-driven spin-conserving
transitions populating the conduction band states in phase two, and emission
of entangled photons into the cavities in phase three. \emph{Lower panel:} Numerically calculated time evolution, in
the coherent limit $\kappa\rightarrow 0$, of the occupation probability of various electronic and
photonic modes. At the end of phase one, marked by
$t_1=\pi/2\Gamma_c$, the electronic occupation per spin in the
heavy-hole band reaches its maximum ($\approx 0.5$). During phase two,
between $t_1$ and $t_2$, the
electronic population is transferred from the heavy-hole band to the conduction band.
Once the conduction band is occupied the electrons can transition from the conduction band to the light-hole band by emitting photons into the cavity. At the end of phase three, marked by $t_3$, the electronic system
is in the product state $\ket{\psi}_{\rm el} \approx
\ket{\lh,\uparrow}_L \ket{\lh,\uparrow}_R$, and the entanglement has
been transferred to the photonic degree of freedom (see text and Fig.~\ref{fig:LogNegPlot}).
\label{fig:CombinedPlot}}
\end{center}
\end{figure}

Once both photons have been
emitted, the electronic singlet has been transferred onto a two-photon
state, ideally given by
\begin{align}
\ket{\psi}_{\rm ph}=\mathcal{N}\left(\ket{\omega_{\updownarrow}^{},\updownarrow}_L^{}\ket{\omega_{\circlearrowleft}^{},\circlearrowleft}_R^{}-\ket{\omega_{\circlearrowleft}^{},\circlearrowleft}_L^{}\ket{\omega_{\updownarrow}^{},\updownarrow}_R^{}\right).\label{eq:25}
\end{align}
Here $\kets{\omega_p,p}$,
with $p\in\{\updownarrow,\circlearrowleft\}$, represents the
photon states emitted into the continuum modes with either linear
($\updownarrow$) or circular ($\circlearrowleft$) polarization, and $\omega_p$ denotes the corresponding transition frequency.
Importantly, because
of the finite linewidth of the electronic levels, the emitted photons are spread out in frequency. Let us
characterize the frequency overlap by
$\varepsilon=1-|\braket{\omega_{\updownarrow}|\omega_{\circlearrowleft}}|^2$. The
normalization of the above photonic state $\ket{\psi}_{\rm ph}$ is
then given
by $\mathcal{N}=(1+\varepsilon)^{-1/2}$.
As
we show next, the entanglement of the state~(\ref{eq:25}) can
be detected by standard polarization measurements, as long as
$\varepsilon$ is below a certain threshold value.

The density matrix of the polarization degree of freedom is obtained
by tracing $\rho_{\rm ph} = \ket{\psi}\bra{\psi}_{\rm ph}$ in Eq.~(\ref{eq:25}) over the frequency degree of freedom,
\begin{align}\label{eq:8}
\rho_{\rm
  pol}&=\frac{1}{1+\varepsilon}\Big[\ket{\updownarrow,\circlearrowleft}\bra{\updownarrow,\circlearrowleft}+\ket{\circlearrowleft,\updownarrow}\bra{\circlearrowleft,\updownarrow}\notag\\
&\phantom{=\frac{1}{1+\varepsilon}\Big(}-(1-\varepsilon)\left(\ket{\updownarrow,\circlearrowleft}\bra{\circlearrowleft,\updownarrow}+\text{h.c.}\right)\Big],
\end{align}
where we have introduced the shorthand notation
$\ket{\updownarrow,\circlearrowleft}\equiv\ket{\updownarrow}_L^{}\otimes\ket{\circlearrowleft}_R^{}$
and similar for the other two-photon polarization states. In the limit $\varepsilon\rightarrow 1$, corresponding to
distinguishable frequencies, the state~(\ref{eq:8}) is separable: $\rho_{\rm
  pol}=\big(\rho_\updownarrow^{(L)}\otimes\rho_\circlearrowleft^{(R)}+\rho_\circlearrowleft^{(L)}\otimes\rho_\updownarrow^{(R)}\big)/2$
with $\rho_p^{(\alpha)}=\ket{p}\bra{p}_\alpha$. In the other limit $\varepsilon\rightarrow 0$, corresponding
to indistinguishable frequencies, the state~(\ref{eq:8}) is maximally entangled:
$\rho_{\rm pol}=\ket{\psi_-}\bra{\psi_-}$ with
$\ket{\psi_-}=\big(\ket{\updownarrow,\leftrightarrow}-\ket{\leftrightarrow,\updownarrow}\big)/\sqrt{2}$. To
obtain the latter expression, we have decomposed the circularly
polarized state as a superposition of two orthogonal linearly
polarized states
$\ket{\circlearrowleft}=(\ket{\updownarrow}+i\ket{\leftrightarrow})/\sqrt{2}$. Thus,
depending on the value of $\varepsilon$, the polarization
degree of freedom may or may
not be entangled.

\begin{figure}[t]
\begin{center}
\includegraphics[scale=0.7, viewport=12 1 238 212]{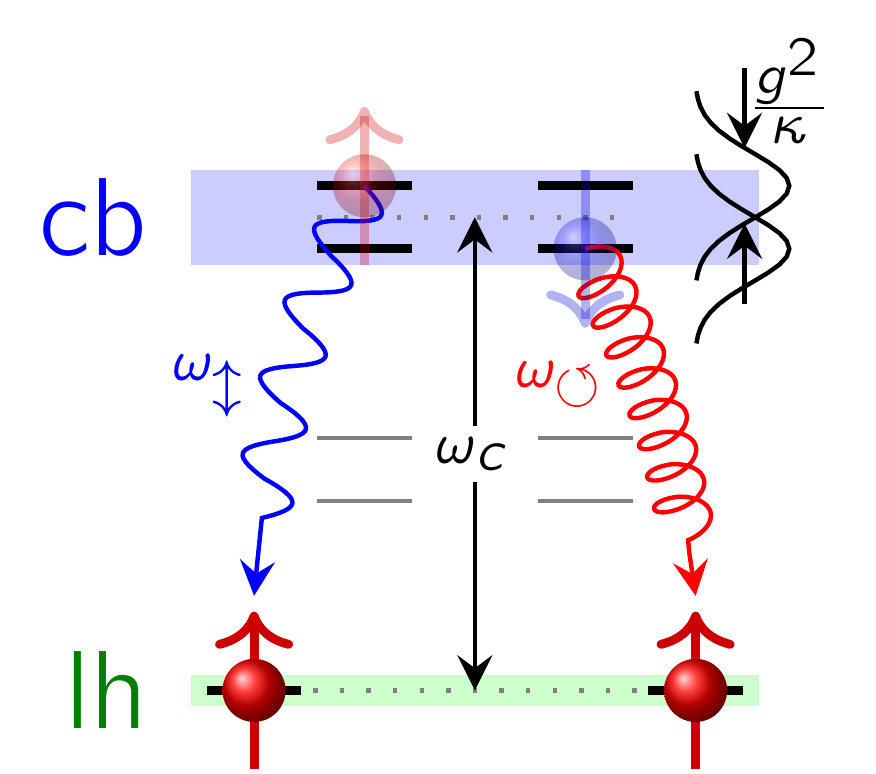}\caption{Schematics of the level scheme
  relevant for the frequency disentangling Purcell emission
  process. We omit the filled lower $\lh$ states because transitions
  to the latter
  are blocked (see text).\label{fig:purcell_levels}}
\end{center}
\end{figure}

\begin{figure*}[t]
\begin{center}
\includegraphics[viewport=0 0 507 137]{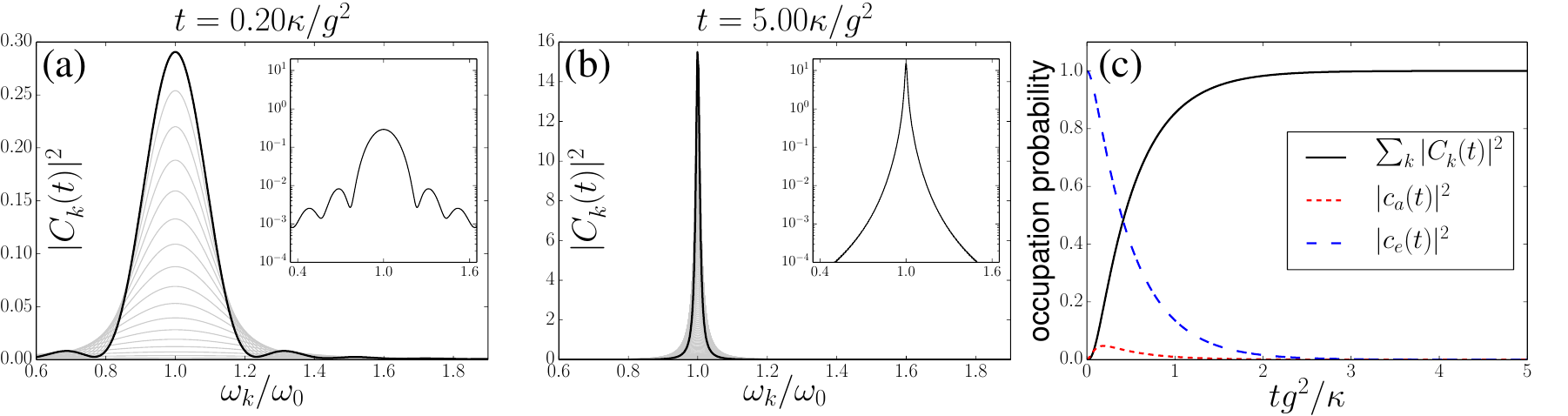}
\caption{\emph{Panels (a) and (b):} Snapshots of the frequency distribution
  of the emitted photon. At short time $t\ll\kappa/g^2$, transient coherent oscillations are clearly
  visible in
  the logarithmic plot shown in the inset (a) and are suppressed at long time
  $t\gg\kappa/g^2$, when the distribution becomes Lorentzian (b). \emph{Panel (c):} Evolution of the occupation probabilities of
  electronic state, cavity mode and integrated continuum modes. Note
  that the cavity population $|c_a(t)|^2$ remains small at all time.\label{fig:purcell_snap}}
\end{center}
\end{figure*}

An experimentally accessible way of demonstrating entanglement in
polarization is provided by the violation of the Clauser-Horne-Shimony-Holt (CHSH) variant of Bell's
inequality~\cite{Clauser-1969a}; by now a standard technique of quantum
optics~\cite{Aspect-1982a}. In our case, we find that $\rho_{\rm pol}$ violates the CHSH
inequality if (see Appendix~\ref{sec:chsh-ineq-entangl})
\begin{align}\label{eq:13}
  \varepsilon < \frac{\sqrt{2}-1}{\sqrt{2}+1}.
\end{align}
To relate $\varepsilon$ with the parameters of our model, we use
the Weisskopf-Wigner (WW) theory~\cite{Weisskopf-1930a} of
the Purcell effect, which allows us to derive analytically
the state of the photons emitted into the continuum by the electronic
system via the cavity.

The relevant part of the level scheme is depicted in Fig.~\ref{fig:purcell_levels}. We consider
the zero-temperature limit where the cavity is initially empty. In each QD, the problem then separates
into two independent Purcell emission processes,
corresponding to transitions from the conduction band levels $\ket{\cb,\uparrow}$ and $\ket{\cb,\downarrow}$
into the unoccupied light-hole state $\ket{\lh,\uparrow}$ (blue and red arrows in Fig.~\ref{fig:purcell_levels}).
For each of these transitions, the photon
emission process can be described
by the Jaynes-Cummings Hamiltonian~\cite{Jaynes-1963a}, where the cavity mode is coupled to a bosonic
quasi-continuum. Within the WW theory, the associated Schr\"odinger
equation can be solved analytically (see Appendix~\ref{supp:WW}) and
the solution is given by
\begin{align}\label{eq:21}
\ket{\psi}=c_e\ket{1,0,\{0\}} + c_a\ket{0,1,\{0\}}+\sum_k C_{k}\ket{0,0,\{1_k\}}.
\end{align}
Here $\ket{n,m,\{s_k\}}$ denotes a state with $n$ electrons in the
conduction band level, $m$ photons in the cavity mode and $s_k$
photons with momentum $k$ in the continuum ($\ket{\{0\}}$ denotes
the vacuum state of the continuum). Since we want to extract the
photons quickly and avoid coherent oscillations between the cavity and
the electrons, we focus on the weak-coupling, near-resonant regime
where $g,|\delta|\ll\kappa$. Here $\delta=\omega_{\rm cav}-\omega_0$ is the detuning of the cavity mode
$\omega_{\rm cav}$ from the spin-conserving and spin-flipping electronic transitions with
frequencies $\omega_0=\omega_{\updownarrow}=(E^{\cb}_{\uparrow}-E^{\lh}_{\uparrow})/\hbar$
and $\omega_0=\omega_{\circlearrowleft}=(E^{\cb}_{\downarrow}-E^{\lh}_{\uparrow})/\hbar$, respectively.

At long time $t\gg \kappa/g^2$, $c_e(t)$ and $c_a(t)$ vanish (see
Fig.~\ref{fig:purcell_snap}, panel (c)), while
the amplitude of the emitted photon $C_{k}(t)$
asymptotically goes towards (see Fig.~\ref{fig:purcell_snap} panels
(a) and (b) and Appendix~\ref{supp:WW})
\begin{align}\label{eq:11}
C_{k}(t\gg
\kappa/g^2)\approx\frac{-\nu_0^{}ge^{-i\omega_k^{}t}}{\left(\kappa_0^{}+i(\omega_0^{}-\omega_k^{})\right)\left(\kappa_c^{}+i(\omega_{\rm
      cav}^{}-\omega_k^{})\right)},
\end{align}
where $\kappa_0^{}\approx g^2/\kappa$, $\kappa_c^{}\approx
\kappa-g^2/\kappa$, $\omega_k^{}$ is the photon frequency in the
continuum, and $\nu_{0}^{}$ is the coupling constant between the
cavity mode and the continuum. The state of the emitted photon can be written as
\begin{align}
\ket{\omega_0}=\sum_kC_{k}(t)\ket{\{1_k\}}.
\end{align}
In the long time limit, the distribution of the
emitted photons is centered on the frequency $\omega_0$ and its width
is determined by the Purcell rate
$\kappa_0^{}$, because this is the smaller of
the two rates $\kappa_c^{}$ and $\kappa_0^{}$.

Equation~(\ref{eq:11}) allows us to evaluate the overlap in frequency
of two photons emitted during the spin-conserving and spin-flipping
transitions. For $\Delta_Z \ll \kappa_0$, we can expand to leading order in $|\omega_{\updownarrow}-\omega_{\circlearrowleft}|/\kappa_0^{}$ and find
\begin{align}\label{eq:14}
\varepsilon=1-\left|\brakets{\omega_{\updownarrow}^{}\big|\omega_{\circlearrowleft}^{}}\right|^2&=1-\Big|\sum_k(C^{\updownarrow}_{k})^*C^{\circlearrowleft}_{k}\Big|^2\approx
\left(\frac{\omega_{\updownarrow}^{}-\omega_{\circlearrowleft}^{}}{2\kappa_0^{}}\right)^2.
\end{align}
Hence, from Eqs.~(\ref{eq:13}) and (\ref{eq:14}), we find that the state of the emitted photons is
entangled in polarization if
\begin{align}\label{eq:16}
\frac{\left|\omega_{\updownarrow}^{}-\omega_{\circlearrowleft}^{}\right|}{g^2/\kappa}<2\left(\sqrt{2}-1\right)\approx 0.83.
\end{align}
Thus, if the linewidth of the conduction band levels induced by the
Purcell effect is larger than the Zeeman
splitting of the conduction band doublet, it is possible to
violate Bell's inequality, thereby demonstrating the
entanglement of the split Cooper pair.

\section{Sensitivity to imperfections}
\begin{figure}[t]
\begin{center}
\includegraphics[viewport=5 2 250 181]{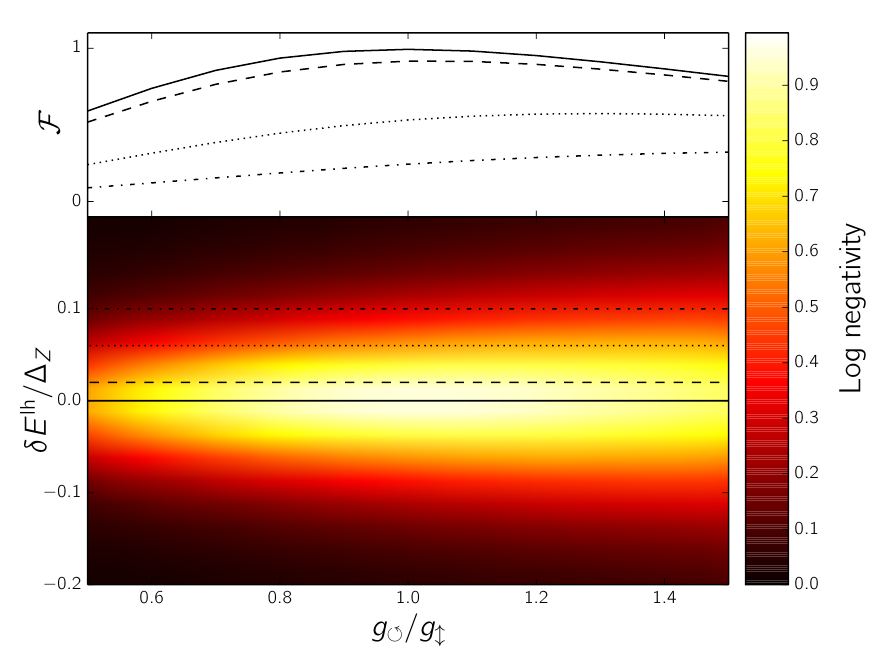}\caption{\emph{Upper panel:} Fidelity $\mathcal{F}=\brakets{\psi|\rho_{\rm
        ph}|\psi}_{\rm ph}$ of the numerically computed photonic
  state $\rho_{\rm ph}$ with the ideal state $\ket{\psi}_{\rm ph}$ of
  Eq.~(\ref{eq:25}). \emph{Lower panel:} Logarithmic negativity of $\rho_{\rm
    ph}$ as a function of photon coupling asymmetry $g_\circlearrowleft/g_\updownarrow$ and detuning $\delta E^{\lh} = \hbar\omega_\updownarrow - (E^{\cb} - E^{\lh})$ between the cavity resonance and the electronic transition frequency. The larger the value of the logarithmic negativity, the more entanglement
  is present. \label{fig:LogNegPlot}}
\end{center}
\end{figure}

So far we have discussed the ideal case without any imperfections. To quantify the sensitivity of the proposed scheme to realistic parameter
variations, we use the numerical simulations for the coherent
system ($\kappa\rightarrow 0$). In this case, the irreversible
Purcell emission in phase three is replaced by coherent Rabi oscillations between the
electronic system and the cavity (see
Fig.~\ref{fig:CombinedPlot}). After half-integer multiples of the Rabi
period, the photonic state in the cavities is ideally given by
Eq.~(\ref{eq:25}). The
fidelity of the actually generated photonic state computed numerically
with this ideal state is shown in the top panel of
Fig.~\ref{fig:LogNegPlot}. To quantify the entanglement of the
photonic state generated in the cavity, we compute its logarithmic
negativity~\cite{Vidal-2002a}. This is shown in the bottom panel of
Fig.~\ref{fig:LogNegPlot}. Both the fidelity and the logarithmic
negativity are shown as a function of the asymmetry of the
electron-photon coupling strengths for the two polarizations
$g_{\circlearrowleft}/g_{\updownarrow}$ and as a
function of the detuning $\delta E^{\lh}$ between the cavity mode and the electronic
transition frequency between conduction band and light-hole band. We stress that since only local
unitary operations are applied to each QD, the positivity of the logarithmic negativity of the
photonic state bears witness to the entanglement of the split Cooper
pair. As expected, the optimal entanglement transfer takes place
closest to resonance and for equal coupling strengths. However,
sizeable and detectable photon entanglement remains even away from the
optimal point. In Appendix~\ref{sec:entangl-pres-electr}, we further show that
photonic entanglement persists even in the presence of finite
electronic decoherence. Finite temperature effects are not included in
the present work but could be an interesting topic for future investigation.

\section{Conclusion}
In conclusion, we have proposed a hybrid electro-optical scheme to detect entanglement of
split Cooper pairs. By mapping the spin entanglement of
electrons to the polarization entanglement of optical photons, we avoid several
difficulties of previous proposals. Provided cavities with little
cross-talk at a
distance smaller than the superconducting coherence length can be
fabricated~\cite{Brossard-2013a} (see also Appendix~\ref{sec:effect-cavity-cross}),
our scheme could be implemented by combining state-of-the-art technologies: a photonic crystal cavity with
coupling strength $g\approx 20 \text{ GHz}$ \cite{Yoshie-2004a} and
linewidth $\kappa \approx 100 \text{ GHz}$ leads to $g^2/\kappa
\approx 4 \text{ GHz}$. This exceeds the Zeeman splitting $\Delta_Z
\approx  0.1 \text{ GHz}$ corresponding to a magnetic field of $10
\text{ mT}$. With a typical Cooper pair splitting rate of $\Gamma_c
\approx 2 \text{ GHz}$ \cite{Herrmann-2010a}, the entire entanglement
transfer can thus be performed fast compared to typical decoherence
rates, $1/T_2 \approx 0.01 \text{ GHz}$ \cite{Kloeffel-2013}. Our
scheme can thus be used to verify
the entanglement-preserving nature of the Cooper pair splitting
process~\cite{Hofstetter2009}; a crucial step towards realizing a
reliable source of electronic EPR pairs in the solid state.

We would like to acknowledge stimulating
discussions with Christoph Bruder. This work was financially
supported by the Swiss SNF and the NCCR
Quantum Science and Technology. SW acknowledges support from the ITN cQOM. TLS is supported by the National Research Fund, Luxembourg (ATTRACT 7556175).

\begin{appendix}

\section{\label{supp:SW} Effective Hamiltonian for Cooper pair
  splitting}
In this appendix, we present a systematic derivation of the Cooper pair tunneling rate using the Schrieffer-Wolff (SW) transformation. The system under investigation is described
by the Hamiltonian $H=H_{\rm SC}+H_{L}+H_{R}+K_{{\rm SC}-{\rm QD}}$, where
\begin{equation}
 H_{\rm SC}= \sum_{{\bf k}\sigma}E_{{\bf k}}\gamma_{{\bf k}\sigma}^\dagger \gamma_{{\bf k}\sigma}^{}
 \label{eq:hamS}
\end{equation}
describes a conventional BCS superconductor. Here, $\gamma_{{\bf k}\sigma}^{}$ is the quasiparticle annihilation operator (with momentum $\bf k$ and spin $\sigma\in\{\uparrow, \downarrow\}$), which is defined by $\gamma_{{\bf k}\sigma}^{}\ket{\rm BCS}=0$, where $\ket{\rm BCS}$ denotes the BCS ground state.
The quasiparticle energies are given by $E_{{\bf k}}=\sqrt{\xi_{{\bf k}}^2+\Delta^2}$, where $\Delta$ is the superconducting gap and $\xi_{{\bf k}}$ is the normal-state single-electron energy as measured
from the chemical potential of the superconductor (henceforth set to zero).

Only the electronic states in the heavy-hole ($\hh$) band of the quantum dots are included in the derivation of the Cooper pair tunneling rate. States in the light-hole and conduction bands can be safely ignored due to their larger detuning from the superconductor's chemical potential.
The Hamiltonian describing the quantum dots is
\begin{equation}
 H_{\alpha}=\sum_{\sigma}\left(E^{(\hh)}+\sigma\frac{\Delta_Z}{2}\right)n_{\alpha\sigma}^{(\hh)},
 \label{eq:hamDQD}
\end{equation}
where $\sigma \in\{ \uparrow,\downarrow\} = \{+,-\}$ and $\alpha\in\{L, R\}$ denotes the left and right quantum dots, respectively. Moreover, $n_{\alpha\sigma}^{(\hh)}=c_{\alpha\sigma}^{(\hh)\dagger}c_{\alpha\sigma}^{(\hh)}$ is the corresponding
number operator given in terms of the electronic creation
($c_{\alpha\sigma}^{(\hh)\dagger}$) and annihilation
($c_{\alpha\sigma}^{(\hh)}$) operators for the electrons in the heavy-hole bands.
For simplicity, we have assumed that the left and right quantum dots have the same orbital and Zeeman energies. Furthermore,
the quantum dots are assumed to be in the Coulomb blockade regime, so double occupancy of the heavy-hole bands is forbidden. The tunnel coupling between the superconductor and the quantum dots is described by
\begin{equation}
 K_{{\rm SC}-{\rm QD}}=\sum_{\sigma}\sum_{\alpha}w_\alpha\left(c_{\alpha\sigma}^{(\hh)}\psi_{\sigma}^{\dagger}({\bf r}_{\alpha}) + \text{h.c.}\right),
 \label{eq:hamDQDtun}
\end{equation}
where $w_{\alpha}$ is the corresponding electron tunneling amplitude and $\psi_{\sigma}^{\dagger}({\bf r}_{\alpha})$ creates an electron (with spin $\sigma$) at position ${\bf r}_{\alpha}$ in the superconductor. Going to momentum space, we can express the tunnel coupling in terms of the electron creation ($d_{{\bf k}\sigma}^{\dagger}$) and annihilation ($d_{{\bf k}\sigma}$) operators for the superconductor, which are related to the quasiparticle operators via the Bogoliubov transformation
 \begin{eqnarray}
  d_{{\bf k}\uparrow}&=&u_{{\bf k}}\gamma_{{\bf k}\uparrow}^{}+v_{{\bf k}}\gamma_{-{\bf k}\downarrow}^{\dagger}, \nonumber \\
  d_{-{\bf k}\downarrow}&=&u_{{\bf k}}\gamma_{-{\bf k}\downarrow}^{}-v_{{\bf k}}\gamma_{{\bf k}\uparrow}^{\dagger}. \label{eq:bogoliubov}
 \end{eqnarray}
Here, $u_{{\bf k}}=(1/\sqrt{2})\sqrt{1+\xi_{{\bf k}}/E_{{\bf k}}}$ and $v_{{\bf k}}=(1/\sqrt{2})\sqrt{1-\xi_{{\bf k}}/E_{{\bf k}}}$ are the usual BCS coefficients.

We express the original Hamiltonian as $H=H_{0}+K_{\rm{SC}- \rm {QD}}$.
We wish to determine a unitary transformation $\mathcal{U}=e^{-S}$ that eliminates $K_{{\rm SC}-{\rm QD}}$ to linear order in $w_{\alpha}$. Choosing the anti-Hermitian operator $S\sim\mathcal{O}(w_{\alpha})$ such that
\begin{equation}
[H_{0}, S]=-K_{{\rm SC}-{\rm QD}},\label{eq:S7}
\end{equation}
the transformed Hamiltonian becomes, to second order in $w_{\alpha}/\Delta$
\begin{equation}
H_{\rm SW}=e^{-S}He^{S}\approx H_{0} + \frac{1}{2}[K_{{\rm SC}-{\rm QD}}, S] +\mathcal{O}(w_{\alpha}^3).
\label{eq:formalsw}
\end{equation}
The solution of~(\ref{eq:S7}) is given by
\begin{equation}
S=\sum_{\alpha}\sum_{{\bf k}\sigma}\left(\gamma_{{\bf k}\sigma}X^{\alpha}_{{\bf k}\sigma}-\text{h.c.}\right),
\end{equation}
where
\begin{eqnarray}
X^{\alpha}_{{\bf k}\uparrow}&=&w_{\alpha}\left(\frac{u_{\bf k}e^{i{\bf k}\cdot{\bf r}_{\alpha}}c^{(\hh)\dagger}_{\alpha\uparrow}}{E_{{\bf k}}-E^{(\hh)} - \frac{\Delta_Z}{2}}+
\frac{v_{\bf k}^{\ast}e^{i{\bf k}\cdot{\bf r}_{\alpha}}c^{(\hh)}_{\alpha\downarrow}}{E_{{\bf k}}+E^{(\hh)} - \frac{\Delta_Z}{2}}\right) \nonumber \\
X^{\alpha}_{{\bf k}\downarrow}&=&w_{\alpha}\left(\frac{u_{\bf k}e^{i{\bf k}\cdot{\bf r}_{\alpha}}c^{(\hh)\dagger}_{\alpha\downarrow}}{E_{{\bf k}}-E^{(\hh)} + \frac{\Delta_Z}{2}}-
\frac{v_{\bf k}^{\ast}e^{i{\bf k}\cdot{\bf r}_{\alpha}}c^{(\hh)}_{\alpha\uparrow}}{E_{{\bf k}}+E^{(\hh)} + \frac{\Delta_Z}{2}}\right).
\label{eq:xtran}
\end{eqnarray}
The effective Hamiltonian at low temperatures and for large Coulomb repulsion is then obtained by projecting
$H_{\rm SW}$ onto the subspace where all quasiparticle states are empty and the two heavy-hole states of a given quantum dot contain at
most one electron.
To second order in $w_{\alpha}$, we obtain
\begin{widetext}
\begin{eqnarray}
H_{\rm eff}&=& H_0 + \sum_{{\bf k}}\Bigg[\frac{w_Lw_R}{2}\Big(\sum_{\sigma, \sigma^{\prime}}
\frac{u_{{\bf k}}v_{{\bf k}}}{E_{{\bf k}}+\sigma E^{(\hh)}+\sigma^{\prime}\Delta_Z/2}\Big)\Big(e^{\sigma\sigma^\prime i {\bf k}\cdot{\delta {\bf r}}} c_{L\downarrow}^{(\hh)\dagger} c_{R\uparrow}^{(\hh)\dagger}
-e^{-\sigma\sigma^\prime i {\bf k}\cdot{\delta {\bf r}}} c_{L\uparrow}^{(\hh)\dagger} c_{R\downarrow}^{(\hh)\dagger}\Big) \nonumber \\
&+&w_Lw_R\Big(\frac{|v_{{\bf k}}|^2}{E_{{\bf k}}-E^{(\hh)}-\Delta_Z/2}
-\frac{|u_{{\bf k}}|^2}{E_{{\bf k}}+E^{(\hh)}+\Delta_Z/2}\Big)c_{L\downarrow}^{(\hh)\dagger}c^{(\hh)}_{R\downarrow}e^{i{\bf k}\cdot\delta {\bf r}} \nonumber \\
&+&w_Lw_R\Big(\frac{|v_{{\bf k}}|^2}{E_{{\bf k}}-E^{(\hh)}+\Delta_Z/2}
-\frac{|u_{{\bf k}}|^2}{E_{{\bf k}}+E^{(\hh)}-\Delta_Z/2}\Big)c_{L\uparrow}^{(\hh)\dagger}c_{R\uparrow}^{(\hh)}e^{i{\bf k}\cdot\delta {\bf r}} + \text{h.c.} \Bigg],
\end{eqnarray}
\end{widetext}
where $\delta {\bf r}={\bf r}_L-{\bf
  r}_R$. The first term in the brackets describes the coherent Cooper pair
splitting while the second and third terms describe an effective spin-conserving
inter-dot coupling. We note that the latter two terms are suppressed by a small factor
$\Delta_Z/\Delta \ll 1$ compared to the first one, and therefore can be safely ignored.
The sum over ${\bf k}$ can be performed by linearizing the spectrum around the Fermi energy and using $u_{{\bf k}}v_{{\bf k}}=\Delta/(2E_{{\bf k}})$.
The effective Hamiltonian can then be written as
\begin{equation}
H_{\rm eff}=H_0+\sum_{\alpha=L,R} \left( \hbar\Gamma_c c^{\dag}_{\alpha\uparrow} c^{\dag}_{\alphabar\downarrow}
+\text{h.c.} \right),
\end{equation}
where
\begin{equation}
\hbar\Gamma_{c}=w_Lw_R\pi\rho_0 \frac{\sin(k_F|\delta{\bf r}|)}{2 k_F|\delta{\bf r}|} \sum_{j = \pm} \frac{e^{-\eta_j |\delta{\bf r}|/(\pi \xi)}}{\eta_j}.
\label{eq:gc}
\end{equation}
Here, $k_F$ is the Fermi momentum, $\xi$ is the superconducting
coherence length, $\rho_0$ is the normal-state density of states
at the chemical potential of the superconductor, and
\begin{align}
    \eta_\pm &= \sqrt{1-\left(\frac{E^{(\hh)} \pm \Delta_Z/2}{\Delta}\right)^2}
\end{align}
On resonance, i.e., for
$E^{(\hh)}=0$ and in the limit
$\delta{\bf r}/\xi\rightarrow 0$, Eq.~(\ref{eq:gc}) reduces to the
expression given in the main text.

\section{Numerical simulation\label{sec:numerical-simulation}}

To describe the dynamics of our entanglement transfer scheme,
we use a real-time simulation of the system from the initial
emission of the Cooper pair into the quantum dots to the final emission of the
polarization entangled photons into the cavities. We will distinguish three phases:

In phase one, we use the gates to load a singlet into the heavy-hole state of the quantum dots. This phase is described by the Hamiltonian, $H_1(t) = H_L + H_R + H_{\rm prox} + H_{\rm dip}(t)$, where
(for $\alpha \in \{ L,R \}$, $\sigma \in \{ \uparrow,\downarrow \} = \{ +,- \}$, and $\nu \in \{\cb, \hh, \lh\}$),
\begin{align}
    H_{\alpha} &= \sum_{\sigma} E^{(\nu)}_{\sigma} n^{(\nu)}_{\alpha\sigma} + U n_\alpha (n_\alpha - 1),\\
    H_{\rm prox} &= \hbar\Gamma_c \sum_{\alpha} \left( c^{(\hh)\dag}_{\alpha\uparrow} c^{(\hh)\dag}_{\alphabar\downarrow} +\text{h.c.} \right), \notag \\
    H_{\rm dip}(t) &= f(t) \left( n_{L} + n_{R} \right).\notag
\end{align}
The left and right dots are described by the Hamiltonians
$H_{\alpha}$, which contain the different Zeeman-split orbital
energies, $E^{(\nu)}_{\sigma} = E^{(\nu)} + \sigma \Delta_Z/2$, and the
charging energy $U$. The electronic creation and annihilation
operators for the individual orbitals are denoted by
$c^{(\nu)\dag}_{\alpha\sigma}$ and $c^{(\nu)}_{\alpha\sigma}$,
respectively. The corresponding number operators are
$n^{(\nu)}_{\alpha\sigma} =
c^{(\nu)\dag}_{\alpha\sigma}c^{(\nu)}_{\alpha\sigma}$ and the total
number of particles on a given dot is denoted by $n_\alpha =
\sum_{\nu\sigma} n^{(\nu)}_{\alpha\sigma}$. The amplitude of the
proximity coupling $\Gamma_c$ can be found using a Schrieffer-Wolff
transformation, see Eq.~(\ref{eq:gc}). It is the dominant coupling
mechanism near resonance, i.e., for $E_{\uparrow,L}^{(\hh)} +
E_{\downarrow,R}^{(\hh)} = E_{\downarrow,L}^{(\hh)} +
E_{\uparrow,R}^{(\hh)} = 0$, because all other possible tunneling
terms between the superconductor and the quantum dots are strongly
suppressed for large $U$ or $\Delta$. The Hamiltonian $H_{\rm dip}(t)$
describes a time-dependent shift of the onsite energies, and will be
used to establish the resonance condition for half a Rabi period,
$f(t) \approx -E^{(\hh)}\Theta(t) \Theta(t_1 - t)$, where $\Theta(t)$ denotes
the Heaviside function. At time $t_1=\pi/(2\Gamma_c)$, there is a high probability that a singlet occupies the quantum dots.

In phase two, the singlet state is pumped from the heavy-hole band
into the conduction band, and in phase three, the conduction band
electrons transition to the light-hole band emitting photons. These phases are governed by the Hamiltonian $H_{2,3} = \sum_{\alpha} ( H_\alpha + H_{\alpha,\rm ph} + H_{\alpha,\rm transfer} + H_{\alpha,\rm drive} )$, where
\begin{align}
    H_{\alpha,\rm ph} &= \sum_{p = \updownarrow,\circlearrowleft} \hbar \omega_{\alpha p}^{} a^\dag_{\alpha p} a_{\alpha p}, \\
    H_{\alpha,\rm drive}(t) &= A_{\rm drive}(t) e^{-i \omega_{\rm{drive}} t} \sum_{\sigma} c^{(\cb)\dag}_{\alpha \sigma} c^{(\hh)}_{\alpha \sigma} + \text{h.c.}, \notag \\
    H_{\alpha,\rm transfer} &= \hbar g \sum_{\sigma} \left[ a_{\alpha \updownarrow}^{} c^{(\cb)\dag}_{\alpha \sigma} c^{(\lh)}_{\alpha \sigma} + a_{\alpha \circlearrowleft}^{} c^{(\cb)\dag}_{\alpha \sigma} c^{(\lh)}_{\alpha, \bar{\sigma}} + \text{h.c.}\right]. \notag
\end{align}
For the numerical simulation, we use two optical cavity modes with
linear and circular polarizations and frequencies
$\omega_\updownarrow$ and $\omega_\circlearrowleft$, respectively, to
simulate the effect of a single cavity mode with a nonzero
linewidth. The cavity modes are described by $H_{\alpha,\rm ph}$. The drive
Hamiltonians $H_{\alpha,\rm drive}(t)$ model the effect of a drive
laser with frequency $\hbar \omega_{\rm{drive}} \approx E^{(\cb)} - E^{(\hh)}$ and
causes spin-conserving Rabi oscillations between the heavy hole and
conduction band. We assume that its amplitude has the form $A_{\rm
  drive}(t) = \hbar A_0\Theta(t - t_1) \Theta(t_2 - t)$, where
$t_2-t_1\approx \pi/(2A_0)$ is about half a Rabi period. Note that in
order for the drive to efficiently transfer both spin states of the
heavy-hole doublet, the width of its frequency spectrum $\sim A_0$ should be larger than
the detuning due to different Zeeman splittings in the heavy hole and
conduction bands. Alternatively one may use two narrow bandwidth
lasers tuned on resonance with each transition. At the end of phase two (at $t = t_2$), the singlet state
will then reside in the conduction band.

Once the conduction band is occupied, the electrons can transition from the conduction band to the light-hole band by
emitting photons into the cavity. This is described by the coupling
Hamiltonian $H_{\alpha,\rm transfer}$, which leads to Rabi oscillations
between the electrons and the cavity photons. In this process, the
electron may (or may not) flip its spin, thereby emitting a circularly
(linearly) polarized photon. Importantly, we assume that the gate
voltages ensure that the lowest heavy-hole state at energy
$E^{(\lh)}_{\downarrow}$ is always occupied, so that transitions into this state are
blocked due to Pauli exclusion principle. For the numerical simulation, we assume that the photon
frequencies are close to resonance with the respective transitions, i.e.,
$\hbar\omega_{\updownarrow} \approx E^{(\cb)}_{\uparrow} -
E^{(\lh)}_{\uparrow}$ and $\hbar\omega_{\circlearrowleft}\approx
E^{(\cb)}_{\downarrow}-E^{(\lh)}_{\uparrow}$. Again, after half a Rabi
period $~\pi/(2g)$ (at time $t = t_3$), ideally the electronic system
is in the product state $\ket{\psi}_{\rm el} \approx \ket{\lh,\uparrow}_L \ket{\lh,\uparrow}_R$,
whereas the photon degree of freedom should now be entangled.

A plot of the numerical result is shown in Fig.~\ref{fig:LogNegPlot} of the main text. It shows the
transfer of electron population between the heavy-hole band at the beginning ($t=t_1$) and
the light-hole band at the end ($t=t_3$) for an optimal choice of drive durations and strengths.
Moreover, it shows an increase in the photon occupation of the
cavities, which are assumed to be empty before the beginning
($t=t_1$), towards the end of phase three ($t=t_3$).
Using these numerical results, it is convenient to quantify the entanglement in the final state
by calculating the logarithmic negativity of the photon state. The logarithmic negativity
is given by $E_{N}(\rho_{\rm{ph}}) = \log_{2}(|| \rho_{\rm{ph}}^{T_{L,R}} ||_1)$,
where $\rho_{\rm{ph}} = {\rm tr}_{\rm{el}} \left[ \rho_{\rm{tot}} \right]$ denotes the
density matrix of photons, and $T_{L,R}$ means partial transposition with respect
to either subsystem $L$ or $R$.
We investigated the logarithmic negativity in the final photon state as a function of 
the ratios $g_{\circlearrowleft}/g_{\updownarrow}$ and $\delta E^{\lh}$ (see Fig.~\ref{fig:LogNegPlot} of the main text).

Let us stress that the entanglement witnessed by the positive values of the
logarithmic negativity can only stem from the entanglement of the
split Cooper pair, since only local unitary operations are performed on
the two subsystems.

\section{CHSH inequality and entanglement of $\rho_{\rm pol}$\label{sec:chsh-ineq-entangl}}
The CHSH variant of Bell's
inequality used in this work to demonstrate entanglement is expressed
in terms of the photon polarization correlation function
\begin{align}\label{eq:1s}
B = {\rm tr}\left[\rho_{\rm pol}\left(L\otimes (R-R') + L'\otimes (R+R')\right)\right].
\end{align}
An appropriate choice for the operators $L$, $L'$, $R$ and $R'$ is
conveniently given by
\begin{align}
L &= Z,\\
L'&=X,\\
R&=\cos(\theta)Z+\sin(\theta)X,\\
R'&=-\sin(\theta)Z+\cos(\theta)X,
\end{align}
in terms of the pseudo-Pauli operators
\begin{align}
Z&=\ket{\updownarrow}\bra{\updownarrow}-\ket{\leftrightarrow}\bra{\leftrightarrow},\\
X&=\ket{\updownarrow}\bra{\leftrightarrow}+\ket{\leftrightarrow}\bra{\updownarrow}.
\end{align}
Here $\theta/2$ is the relative angle between the polarizing beam
splitter settings of the left and right observers. A state is entangled
in polarization if
\begin{align}\label{eq:15s}
|B|>2,\quad\text{for some } \theta.
\end{align}
In our case we find with Eq.~(\ref{eq:8}) of the main text that
\begin{align}
|B|&=2\frac{1-\varepsilon}{1+\varepsilon}\left|\cos(\theta)+\sin(\theta)\right|.
\end{align}
Maximizing $|B|$ over $\theta$ yields $\theta=\pi/4$ and the condition
for
entanglement of $\rho_{\rm pol}$ given by Eq.~(\ref{eq:13}) of the main text.

\section{\label{supp:WW} Weisskopf-Wigner theory of the Purcell effect
}
In this appendix we derive analytically the amplitudes $c_e(t)$,
$c_a(t)$ and $C_{k}(t)$ of the
Weisskopf-Wigner (WW) Ansatz of Eq.~(\ref{eq:21}) of the main text. Since the
left and right subsystems evolve independently at this stage we suppress the $\alpha = L,R$ index, and focus only on one side of the
system. The photon pair state is then immediately obtained by
linearity. Our starting
point is the Hamiltonian (we set $\hbar=1$)
\begin{align}
H &= H_{\rm JC}+H_{\rm bath}+K\label{eq:17s},
\\
H_{\rm JC}&=
\omega_c^{}a^{\dagger}a+\frac{\omega_0^{}}{2}\sigma^z+ g\left(a\sigma^++a^{\dagger}\sigma^-\right)\label{eq:18s},\\
H_{\rm bath}&=\sum_k\omega_k^{}b_k^{\dagger}b_k^{}\label{eq:19s},\\
K&=\nu_0^{}\sum_k\left(b_k^{}a^{\dagger}+b_k^{\dagger}a\right).\label{eq:20s}
\end{align}
For the spin-conserving transition with transition frequency $\omega_0=\omega_{\updownarrow}\equiv E_{\uparrow}^{({\cb})}-E_{\uparrow}^{({\lh})}$, $\sigma^z$ represents the
Pauli matrix for the effective two-level system consisting of
spin-$\uparrow$ conduction band level and the spin-$\uparrow$ light-hole state, i.e.,
\begin{align}
    \sigma_z &= c^{(\cb)^\dag}_\uparrow c^{(\cb)}_\uparrow - c^{(\lh)^\dag}_\uparrow c^{(\lh)}_\uparrow, \notag \\
    \sigma^+ &= c^{(\cb)^\dag}_\uparrow c^{(\lh)}_\uparrow, \notag \\
    \sigma^- &= [\sigma^+]^\dag.
\end{align}
For the spin-flipping transition with frequency
$\omega_{\circlearrowleft}\equiv E_{\downarrow}^{(\cb)}-E_{\uparrow}^{(\lh)}$, the Pauli matrices are defined analogously. Next, $a$ ($a^{\dagger}$) represents the annihilation (creation)
operator for a photon in the cavity mode with frequency
$\omega_c$. The coupling strengths between the electronic
transition and the cavity mode is denoted with $g$, and $b_k^{}$ ($b_k^{\dagger}$) denotes the annihilation (creation) operator
for a photon with frequency $\omega_k$ in the continuum. In the
wide-band limit, the coupling strength $\nu_0^{}$ between the one dimensional quasi-continuum and the cavity
mode determines the cavity linewidth as $\kappa = \nu_0^2L/(2c)$
where $L/(2\pi)$ is the quasi-continuum mode volume and $c$ the velocity of
light.

Substituting the ansatz of Eq.~(\ref{eq:21}) in the main text into the
associated Schr\"odinger equation yields the differential equations
\begin{align}
i\frac{d}{dt}\begin{pmatrix}c_e(t)\\c_a(t)\\C_{k}(t)\end{pmatrix}=
\begin{pmatrix}
    \omega_0^{}c_e(t)+g c_a(t)\\
    gc_e(t)+\omega_c^{}c_a(t)+\nu_0^{}\sum_k C_{k}(t)\\
    \nu_kc_a(t)+\omega_k^{}C_{k}(t)
\end{pmatrix}.
\end{align}
The solution is most easily obtained by Laplace transform using the
initial conditions $c_e(0)=1$,
$c_a(0)=C_{k}(0)=0$. In Laplace space, we then find
the following algebraic equations ($s$ denotes the Laplace variable)
\begin{align}
sc_e(s)-1&=-i\omega_0^{}c_e(s)-igc_a(s),\label{eq:12b}\\
sc_a(s)&=-igc_e(s)-i\omega_c^{}c_a(s)-i\nu_0^{}\sum_k
C_{k}(s)\label{eq:13b},\\
sC_{k}(s)&=-i\nu_0^{}c_a(s)-i\omega_k^{}C_{k}(s).\label{eq:14b}
\end{align}
Solving for the Laplace amplitudes we find
\begin{align}
c_e(s)&=\frac{s+i\omega_c^{}+\kappa}{(s+i\omega_0^{})(s+i\omega_c^{})+(s+i\omega_0^{})\kappa+g^2}\label{eq:22s},\\
c_a(s)&=\frac{-ig}{(s+i\omega_0^{})(s+i\omega_c^{})+(s+i\omega_0^{})\kappa+g^2}\label{eq:23s},\\
C_{k}(s)&=\frac{-\nu_0g}{s+i\omega_k^{}}\frac{1}{(s+i\omega_0^{})(s+i\omega_c^{})+(s+i\omega_0^{})\kappa+g^2}.
\end{align}
Here, we have applied the WW approximation and
introduced the cavity damping rate $\kappa$ according to
\begin{align}
\sum_k\frac{\nu_0^2}{s+i\omega_k^{}}\approx\kappa+i\Delta\Omega.
\end{align}
The imaginary part $\Delta\Omega$ yields a frequency renormalization similar to the
Lamb-shift, which we shall ignore in the following, as this shift is
typically small in the optical frequency regime.
The WW approximation is valid for weak enough damping, such that
$\kappa\ll\omega_c,\omega_0$ and is essentially equivalent to a
Born-Markov approximation as we have established by comparing the analytic
results below for the intra-cavity and electronic states with a
numerical Lindblad master equation calculation (not shown). Note that in the
optical regime, the above condition is easily satisfied and the
WW approximation is expected to be adequate.

The poles of $c_e(s)$ and $c_a(s)$ are found to be given by
\begin{align}
s_{\pm} =
-i\frac{\omega_0^{}+\omega_c^{}}{2}-\frac{\kappa}{2}\pm\frac{1}{2}\sqrt{\kappa^2-\delta^2-4g^2+2i\kappa\delta},
\end{align}
where we have defined the detuning $\delta\equiv\omega_c-\omega_0$.
$C_{k}(s)$ has an additional imaginary pole at
$s_0=-i\omega_k$. In the regime of interest $\delta,g\ll\kappa$, the
poles are well approximated to order $(g/\kappa)^2$ and
$(\delta/\kappa)^2$ by
\begin{align}
s_{+}&\approx -i\left(\omega_0^{}-\delta
  \left(\frac{g}{\kappa}\right)^2\right)-\frac{g^2}{\kappa},\\
s_{-}&\approx -i\left(\omega_c^{}+\delta
  \left(\frac{g}{\kappa}\right)^2\right)-\kappa +\frac{g^2}{\kappa}.
\end{align}
In the optical regime, the frequency shifts may further be safely
neglected since $\omega_0,\omega_c\gg|\delta|$. Hence,
$s_+=-i\omega_0-g^2/\kappa$ and $s_-=-i\omega_c-\kappa+g^2/\kappa$.
The inverse Laplace transform of the amplitudes amounts to a
summation over residues and yields, to second order in
$g/\kappa$ and $\delta/\kappa$
\begin{align}
c_e(t)&=\frac{1}{\Delta_k^c-\Delta_k^0}\left[\left(i\delta+\kappa_c^{}\right)e^{(-i\omega_0^{}-\kappa_0^{})t}-\kappa_0^{}e^{(-i\omega_c^{}-\kappa_c^{})t}\right]\label{eq:2s},\\
c_a(t)&=\frac{-ig}{\Delta_k^c-\Delta_k^0}\left(e^{(-i\omega_0^{}-\kappa_0^{})t}-e^{(-i\omega_c^{}-\kappa_c^{})t}\right)\label{eq:3s},\\
C_{k}(t)&=-\nu_0g\Bigg[\frac{e^{-i\omega_k^{}t}}{\Delta_k^c\Delta_k^0}
+\frac{1}{\Delta_k^c-\Delta_k^0}\left(\frac{e^{(-i\omega_c^{}-\kappa_c^{})t}}{\Delta_k^c}-\frac{e^{(-i\omega_0^{}-\kappa_0^{})t}}{\Delta_k^0}\right)\Bigg]\label{eq:4s},
\end{align}
with
\begin{align}
\Delta^{0}_{k}&=
\kappa_0^{}+i\omega^{}_0-i\omega^{}_k,\quad\kappa_0^{}\approx \frac{g^2}{\kappa}\label{eq:5s},\\
\Delta^{c}_{k}&=
\kappa_c^{}+i\omega^{}_c-i\omega^{}_k,\quad\kappa_c^{}\approx \kappa-\frac{g^2}{\kappa}.\label{eq:6s}
\end{align}
In the long time limit $t\gg \kappa/g^2$ we obtain from
Eq.~(\ref{eq:4s}) the results of Eq.~(\ref{eq:11}) in the main text.
Fig.~\ref{fig:purcell_snap} of the main text illustrates Eqs.~(\ref{eq:2s}) to (\ref{eq:4s})
for the resonant case $\delta=0$.

\begin{figure}[t]
\begin{center}
\includegraphics[width=0.8\columnwidth, viewport=0 0 394 282]{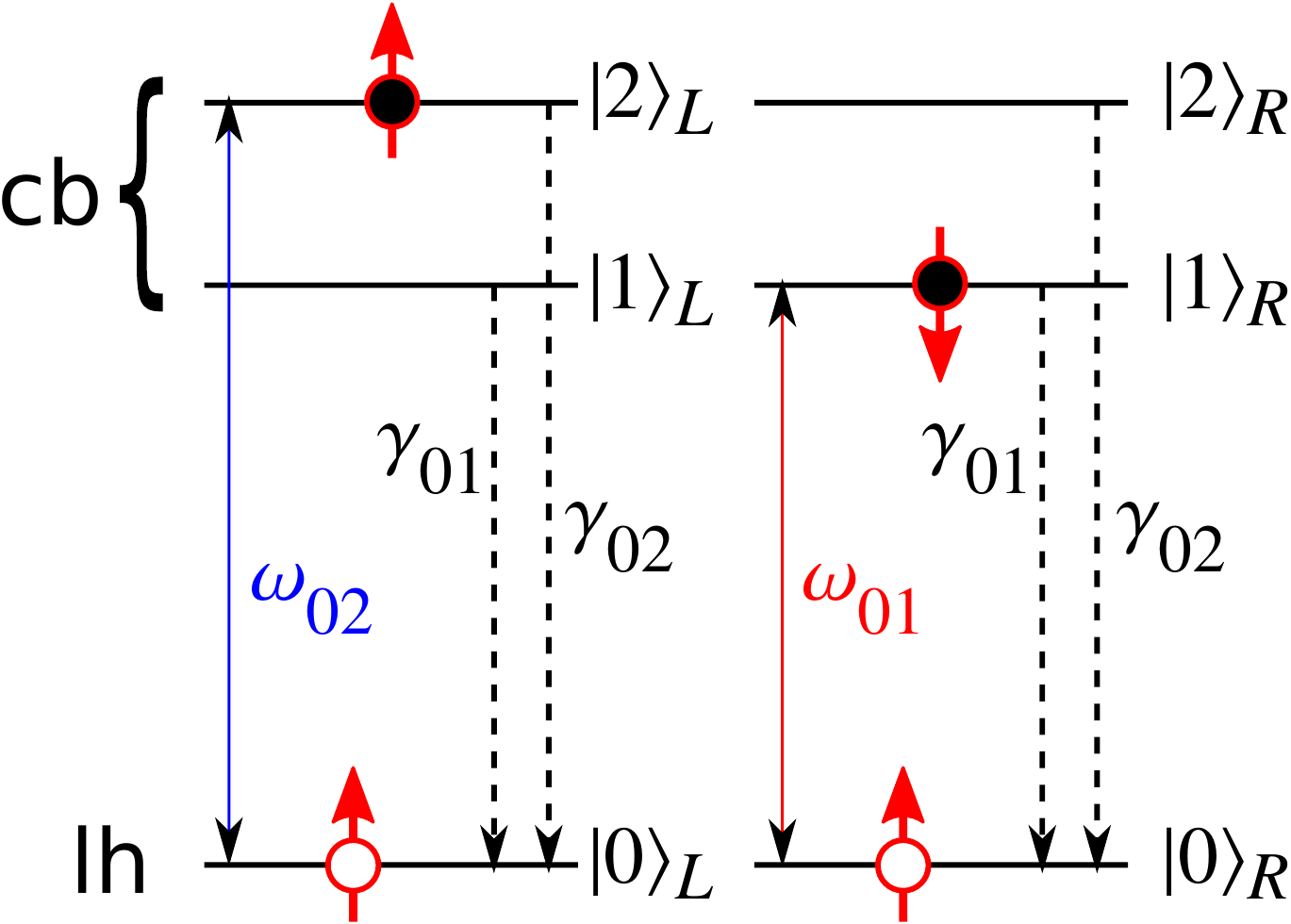}
\caption{Reduced model for the numerical simulation of CHSH inequality
  violation in the presence of electronic decoherence. The dashed
  arrows represent non-radiative electronic decoherence.\label{fig:reduced_model}}
\end{center}
\end{figure}

\section{Entanglement in the presence of electronic decoherence\label{sec:entangl-pres-electr}}
In the main text we have assumed that the photon emission process
takes place on a time scale short compared to the coherence time $T_2$
of the QDs. Here, we investigate numerically the effect of
electronic decoherence on the CHSH inequality violation of the intra-cavity photon state for
a perfect cavity with $\kappa=0$. To this end we consider the reduced
QD model consisting of two three-level ``atoms'', each resonantly coupled to two
cavity modes as depicted in Fig.~\ref{fig:reduced_model}.

\begin{figure}[t]
\begin{center}
\includegraphics[width=\columnwidth]{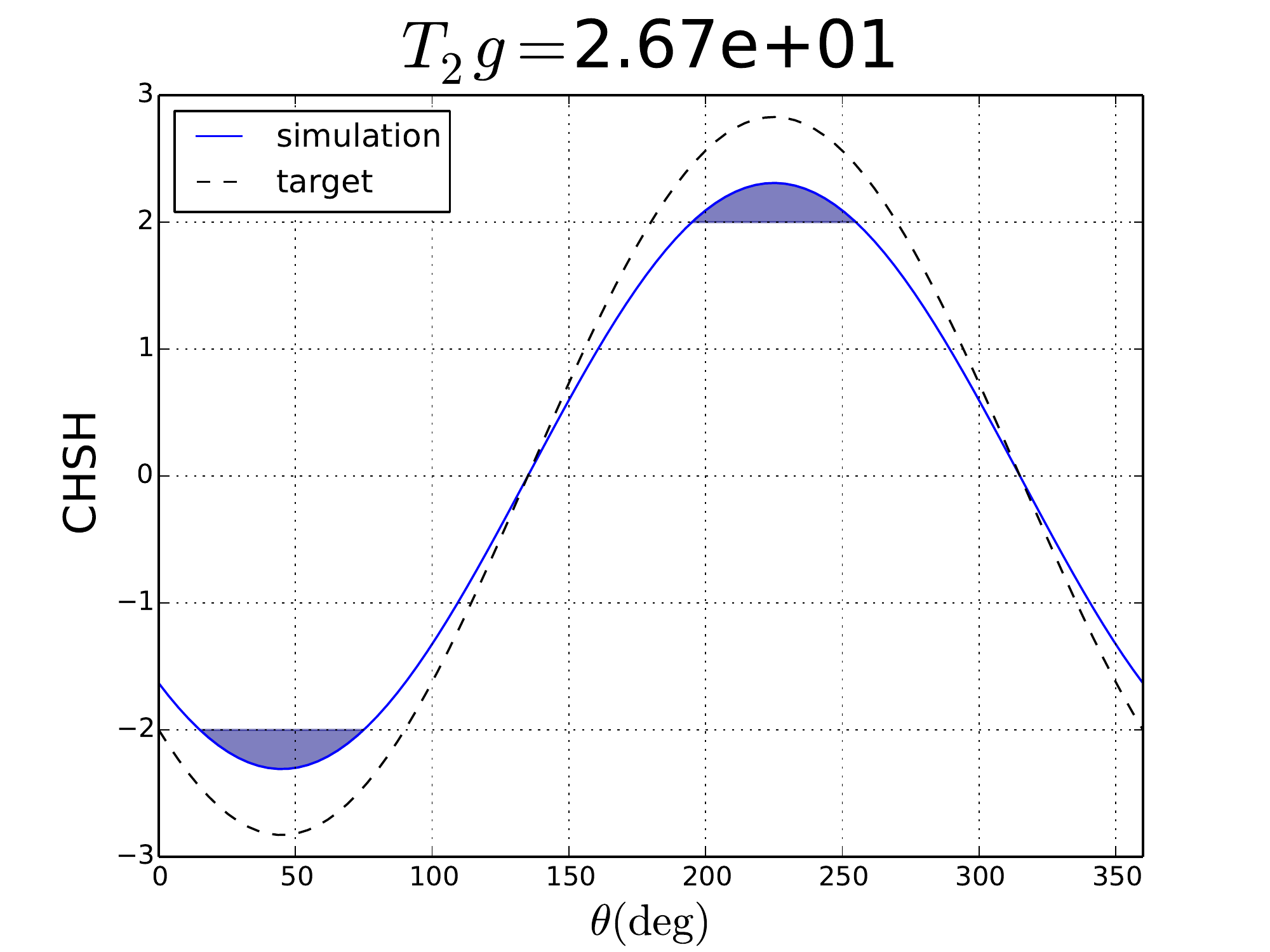}
\caption{Violation of the CHSH inequality with the cavity photons in the presence of
  electronic decoherence. The shaded areas indicate regions where the
  CHSH inequality is violated.\label{fig:Bell_relax}}
\end{center}
\end{figure}

We shall consider electronic
relaxation determined by the non-radiative relaxation rates
$\gamma_{01}$ and $\gamma_{02}$ for transitions from states $\ket{2}$ and
$\ket{1}$ to the state $\ket{0}$ as well as electronic dephasing with
rates $\gamma_{01}^{\varphi}$ and $\gamma_{02}^{\varphi}$ due
to fluctuations of the corresponding transition energies. This corresponds to 
an effective coherence time for the
double dot of $T_2\approx
1/(\gamma_{01}/2+\gamma_{02}/2+\gamma_{01}^{\varphi}+\gamma_{02}^{\varphi})$. As before, the Hamiltonian of both
QD-cavity systems decouples into a sum as $H=H_L+H_R$ with (we set
$\hbar =1$)
\begin{align}
H_{\alpha}&=\omega_{01}\left(\ket{1}\bra{1}+a_{\alpha}^{\dagger}a_{\alpha}^{}\right)+\omega_{02}\left(\ket{2}\bra{2}_{\alpha}+b_{\alpha}^{\dagger}b_{\alpha}^{}\right)\notag\\
&+g\left(\ket{0}\bra{1}_{\alpha}^{}a_{\alpha}^{\dagger}+\ket{0}\bra{2}_{\alpha}b_{\alpha}^{\dagger}+{\rm
    h.c.}\right),\quad\alpha\in\{L,R\}.
\end{align}
Note that, in order to most clearly distinguish the effect of decoherence from
other effects, we choose equal coupling strength $g$ for both
transitions. The evolution of the state of the system
$\rho$, is
described within the Born-Markov approximation by the zero-temperature
Lindblad master equation
\begin{align}
\dot\rho&=-i[H,\rho]+\sum_{\alpha=L,R}\left(\gamma_{01}^{}\mathcal{D}[\sigma_{01}^{\alpha,-}] +\gamma_{02}^{}\mathcal{D}[\sigma_{02}^{\alpha,-}]
\right)\rho\nonumber\\
&+\sum_{\alpha=L,R}\left(\gamma_{01}^{\varphi}\mathcal{D}[\sigma_{01}^{\alpha,z}] +\gamma_{02}^{\varphi}\mathcal{D}[\sigma_{02}^{\alpha,z}]
\right)\rho,
\end{align}
with $\mathcal{D}[O]\rho=\left(2O\rho O^\dagger-O^\dagger O\rho -
  \rho O^\dagger O\right)/2$,
$\sigma_{ij}^{\alpha,-}=\ket{i}\bra{j}_{\alpha}$ and $\sigma_{ij}^{\alpha,z}=\ket{j}\bra{j}_{\alpha}-\ket{i}\bra{i}_{\alpha}$.
The photonic state $\rho_{\rm ph}={\rm tr}_{\rm el}\left[\rho\right]$, is obtained by tracing $\rho$ over the electronic
degrees of freedom.
Fig.~\ref{fig:Bell_relax} shows the CHSH correlation of
Eq.~(\ref{eq:1s}) as a function of the relative angle $\theta$ between
left and right observers, for $T_2g\approx 26.7$. While
decoherence clearly weakens the CHSH correlation, this simulation
essentially demonstrates that as long as the coupling strength $g$ is
large compared with the decoherence rate $T_2^{-1}$ of the QD, the CHSH
inequality can be violated.

\section{Effect of cavity cross-talk\label{sec:effect-cavity-cross}}
In this appendix we investigate quantitatively to what extent cavity
cross-talk affects our entanglement
transfer scheme. To this end we perform a numerical simulation
of the coherent system including a polarization conserving cavity-cavity coupling term of the
form $g_{\rm ph}(a_L^{\dagger}a_R^{} + b_L^{\dagger}b_R^{} +{\rm
  h.c.})$. We use the notation of Appendix~\ref{sec:entangl-pres-electr}.
\begin{figure}[ht]
\includegraphics[width=\columnwidth]{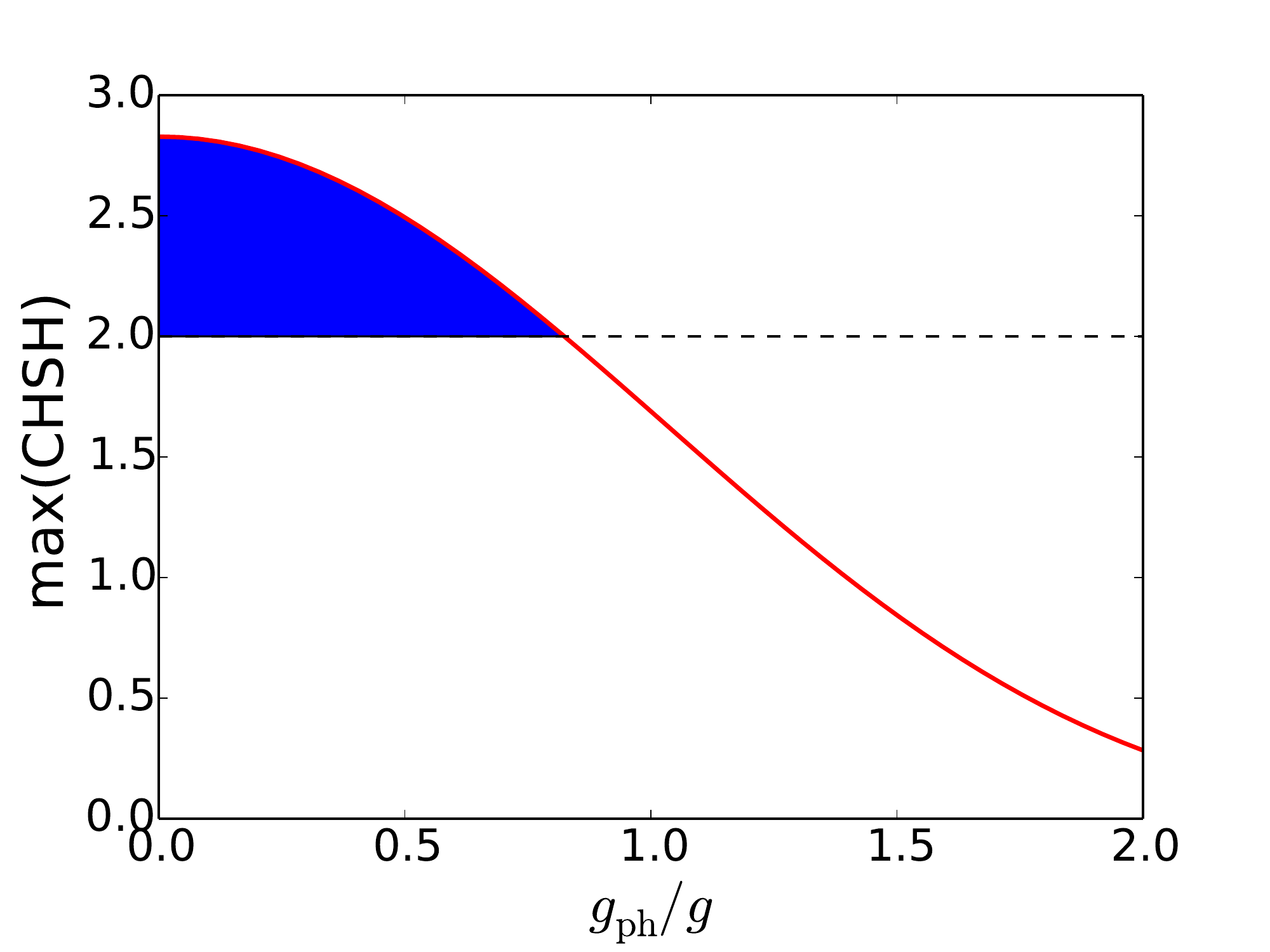}
\caption{Maximal violation of the CHSH inequality after half a
  Rabi period in the presence of
  inter-cavity coupling with strength $g_{\rm ph}$. When the
  inter-cavity coupling is weaker than roughly the electron-photon
  coupling in each cavity, the CHSH inequality is violated (shaded area).\label{fig:inter_cav}}
\end{figure}
Fig.~\ref{fig:inter_cav} shows the maximal Bell inequality violation achievable
after half a (bare) Rabi period $\sim 1/g$. We see that while the correlation
signal is clearly reduced by a finite inter-cavity coupling, it still
surpasses the threshold of $2.0$ demonstrating entanglement, roughly as long
as $g_{\rm ph}<g$. Furthermore, in the experimentally relevant case
where the photons are coupled out of the cavities, the population of
the cavity modes remains small at all times and is of order
$(g/\kappa)^2\ll 1$ (see Appendix.~\ref{supp:WW} and panel (c) of
Fig.~\ref{fig:purcell_snap}). Hence the effective cavity-cavity coupling
rate is reduced compared with the coherent case.
\section{Typical values of the parameters\label{sec:typic-valu-param}}

In this appendix, we estimate the typical range of values for various
parameters used in our theory. A typical value of the coupling strength between the quantum dot and the cavity field is $g\approx 20$ GHz~[\onlinecite{Yoshie-2004a}].
Choosing a cavity decay rate of $\kappa\approx 100$ GHz, leads to an induced bandwidth of $g^2/\kappa\approx 4$ GHz for the conduction band levels. The Zeeman splitting for a magnetic field strength of 10 mT is $\sim 0.1$ GHz. Thus for
these parameters by using a magnetic field which is weaker than 10 mT,
Bell's inequality can be violated. A typical intrinsic coherence time of the self-assembled quantum dots is $T_2\sim 0.1$ $\mu$s ($1/T_2 \sim 0.01$ GHz) [\onlinecite{Kloeffel-2013}].
A typical value of Cooper pair splitting rate is $\Gamma_c\approx 2$ GHz~[\onlinecite{Herrmann-2010a}]. Thus the various conditions ($g^2/\kappa \gg \Delta_Z^{\textrm{(cb)}}$, $g/\kappa \ll 1$,
$g^2/\kappa \gg 1/T_2$, and $\Gamma_c\gg 1/T_2$) essential for
demonstrating entanglement in our scheme, can be satisfied with current technology.

\end{appendix}


\begin{thebibliography}{32}
\expandafter\ifx\csname natexlab\endcsname\relax\def\natexlab#1{#1}\fi
\expandafter\ifx\csname bibnamefont\endcsname\relax
  \def\bibnamefont#1{#1}\fi
\expandafter\ifx\csname bibfnamefont\endcsname\relax
  \def\bibfnamefont#1{#1}\fi
\expandafter\ifx\csname citenamefont\endcsname\relax
  \def\citenamefont#1{#1}\fi
\expandafter\ifx\csname url\endcsname\relax
  \def\url#1{\texttt{#1}}\fi
\expandafter\ifx\csname urlprefix\endcsname\relax\def\urlprefix{URL }\fi
\providecommand{\bibinfo}[2]{#2}
\providecommand{\eprint}[2][]{\url{#2}}

\bibitem[{\citenamefont{Schr{\"o}dinger}(1935)}]{Schroedinger-1935a}
\bibinfo{author}{\bibfnamefont{E.}~\bibnamefont{Schr{\"o}dinger}},
  \bibinfo{journal}{Math. Proc. Cambridge} \textbf{\bibinfo{volume}{31}},
  \bibinfo{pages}{555} (\bibinfo{year}{1935}).

\bibitem[{\citenamefont{Bouwmeester et~al.}(1997)\citenamefont{Bouwmeester,
  Pan, Mattle, Eibl, Weinfurter, and Zeilinger}}]{Bouwmeester-1997a}
\bibinfo{author}{\bibfnamefont{D.}~\bibnamefont{Bouwmeester}},
  \bibinfo{author}{\bibfnamefont{J.-W.} \bibnamefont{Pan}},
  \bibinfo{author}{\bibfnamefont{K.}~\bibnamefont{Mattle}},
  \bibinfo{author}{\bibfnamefont{M.}~\bibnamefont{Eibl}},
  \bibinfo{author}{\bibfnamefont{H.}~\bibnamefont{Weinfurter}},
  \bibnamefont{and}
  \bibinfo{author}{\bibfnamefont{A.}~\bibnamefont{Zeilinger}},
  \bibinfo{journal}{Nature} \textbf{\bibinfo{volume}{390}},
  \bibinfo{pages}{575} (\bibinfo{year}{1997}).

\bibitem[{\citenamefont{Jozsa and Linden}(2003)}]{Jozsa-2003a}
\bibinfo{author}{\bibfnamefont{R.}~\bibnamefont{Jozsa}} \bibnamefont{and}
  \bibinfo{author}{\bibfnamefont{N.}~\bibnamefont{Linden}},
  \bibinfo{journal}{Proc. R. Soc. A} \textbf{\bibinfo{volume}{459}},
  \bibinfo{pages}{2011} (\bibinfo{year}{2003}).

\bibitem[{\citenamefont{Jennewein et~al.}(2000)\citenamefont{Jennewein, Simon,
  Weihs, Weinfurter, and Zeilinger}}]{Jennewein-2000a}
\bibinfo{author}{\bibfnamefont{T.}~\bibnamefont{Jennewein}},
  \bibinfo{author}{\bibfnamefont{C.}~\bibnamefont{Simon}},
  \bibinfo{author}{\bibfnamefont{G.}~\bibnamefont{Weihs}},
  \bibinfo{author}{\bibfnamefont{H.}~\bibnamefont{Weinfurter}},
  \bibnamefont{and}
  \bibinfo{author}{\bibfnamefont{A.}~\bibnamefont{Zeilinger}},
  \bibinfo{journal}{Phys. Rev. Lett.} \textbf{\bibinfo{volume}{84}},
  \bibinfo{pages}{4729} (\bibinfo{year}{2000}).

\bibitem[{\citenamefont{Giovannetti et~al.}(2011)\citenamefont{Giovannetti,
  Lloyd, and Maccone}}]{Giovannetti-2011a}
\bibinfo{author}{\bibfnamefont{V.}~\bibnamefont{Giovannetti}},
  \bibinfo{author}{\bibfnamefont{S.}~\bibnamefont{Lloyd}}, \bibnamefont{and}
  \bibinfo{author}{\bibfnamefont{L.}~\bibnamefont{Maccone}},
  \bibinfo{journal}{Nat. Photonics} \textbf{\bibinfo{volume}{5}},
  \bibinfo{pages}{222} (\bibinfo{year}{2011}).

\bibitem[{\citenamefont{Bell}(1964)}]{Bell-1964a}
\bibinfo{author}{\bibfnamefont{J.~S.} \bibnamefont{Bell}},
  \bibinfo{journal}{Physics} \textbf{\bibinfo{volume}{1}}, \bibinfo{pages}{195}
  (\bibinfo{year}{1964}).

\bibitem[{\citenamefont{Aspect et~al.}(1982)\citenamefont{Aspect, Dalibard, and
  Roger}}]{Aspect-1982a}
\bibinfo{author}{\bibfnamefont{A.}~\bibnamefont{Aspect}},
  \bibinfo{author}{\bibfnamefont{J.}~\bibnamefont{Dalibard}}, \bibnamefont{and}
  \bibinfo{author}{\bibfnamefont{G.}~\bibnamefont{Roger}},
  \bibinfo{journal}{Phys. Rev. Lett.} \textbf{\bibinfo{volume}{49}},
  \bibinfo{pages}{1804} (\bibinfo{year}{1982}).

\bibitem[{\citenamefont{Einstein et~al.}(1935)\citenamefont{Einstein, Podolsky,
  and Rosen}}]{Einstein-1935a}
\bibinfo{author}{\bibfnamefont{A.}~\bibnamefont{Einstein}},
  \bibinfo{author}{\bibfnamefont{B.}~\bibnamefont{Podolsky}}, \bibnamefont{and}
  \bibinfo{author}{\bibfnamefont{N.}~\bibnamefont{Rosen}},
  \bibinfo{journal}{Phys. Rev.} \textbf{\bibinfo{volume}{47}},
  \bibinfo{pages}{777} (\bibinfo{year}{1935}).

\bibitem[{\citenamefont{Bennett and DiVincenzo}(2000)}]{Bennett2000}
\bibinfo{author}{\bibfnamefont{C.~H.} \bibnamefont{Bennett}} \bibnamefont{and}
  \bibinfo{author}{\bibfnamefont{D.~P.} \bibnamefont{DiVincenzo}},
  \bibinfo{journal}{Nature} \textbf{\bibinfo{volume}{404}},
  \bibinfo{pages}{247} (\bibinfo{year}{2000}).

\bibitem[{\citenamefont{Choi et~al.}(2000)\citenamefont{Choi, Bruder, and
  Loss}}]{choi00}
\bibinfo{author}{\bibfnamefont{M.-S.} \bibnamefont{Choi}},
  \bibinfo{author}{\bibfnamefont{C.}~\bibnamefont{Bruder}}, \bibnamefont{and}
  \bibinfo{author}{\bibfnamefont{D.}~\bibnamefont{Loss}},
  \bibinfo{journal}{Phys. Rev. B} \textbf{\bibinfo{volume}{62}},
  \bibinfo{pages}{13569} (\bibinfo{year}{2000}).

\bibitem[{\citenamefont{Recher et~al.}(2001)\citenamefont{Recher, Sukhorukov,
  and Loss}}]{recher01}
\bibinfo{author}{\bibfnamefont{P.}~\bibnamefont{Recher}},
  \bibinfo{author}{\bibfnamefont{E.~V.} \bibnamefont{Sukhorukov}},
  \bibnamefont{and} \bibinfo{author}{\bibfnamefont{D.}~\bibnamefont{Loss}},
  \bibinfo{journal}{Phys. Rev. B} \textbf{\bibinfo{volume}{63}},
  \bibinfo{pages}{165314} (\bibinfo{year}{2001}).

\bibitem[{\citenamefont{Lesovik et~al.}(2001)\citenamefont{Lesovik, Martin, and
  Blatter}}]{Lesovik-2001a}
\bibinfo{author}{\bibfnamefont{G.~B.} \bibnamefont{Lesovik}},
  \bibinfo{author}{\bibfnamefont{T.}~\bibnamefont{Martin}}, \bibnamefont{and}
  \bibinfo{author}{\bibfnamefont{G.}~\bibnamefont{Blatter}},
  \bibinfo{journal}{Eur. Phys. J. B} \textbf{\bibinfo{volume}{24}},
  \bibinfo{pages}{287} (\bibinfo{year}{2001}).

\bibitem[{\citenamefont{Hofstetter et~al.}(2009)\citenamefont{Hofstetter,
  Csonka, Nygard, and Sch\"onenberger}}]{Hofstetter2009}
\bibinfo{author}{\bibfnamefont{L.}~\bibnamefont{Hofstetter}},
  \bibinfo{author}{\bibfnamefont{S.}~\bibnamefont{Csonka}},
  \bibinfo{author}{\bibfnamefont{J.}~\bibnamefont{Nygard}}, \bibnamefont{and}
  \bibinfo{author}{\bibfnamefont{C.}~\bibnamefont{Sch\"onenberger}},
  \bibinfo{journal}{Nature} \textbf{\bibinfo{volume}{461}},
  \bibinfo{pages}{960} (\bibinfo{year}{2009}).

\bibitem[{\citenamefont{Herrmann et~al.}(2010)\citenamefont{Herrmann, Portier,
  Roche, {Levy Yeyati}, Kontos, and Strunk}}]{Herrmann-2010a}
\bibinfo{author}{\bibfnamefont{L.~G.} \bibnamefont{Herrmann}},
  \bibinfo{author}{\bibfnamefont{F.}~\bibnamefont{Portier}},
  \bibinfo{author}{\bibfnamefont{P.}~\bibnamefont{Roche}},
  \bibinfo{author}{\bibfnamefont{A.}~\bibnamefont{{Levy Yeyati}}},
  \bibinfo{author}{\bibfnamefont{T.}~\bibnamefont{Kontos}}, \bibnamefont{and}
  \bibinfo{author}{\bibfnamefont{C.}~\bibnamefont{Strunk}},
  \bibinfo{journal}{Phys. Rev. Lett.} \textbf{\bibinfo{volume}{104}},
  \bibinfo{pages}{026801} (\bibinfo{year}{2010}).

\bibitem[{\citenamefont{Das et~al.}(2012)\citenamefont{Das, Ronen, Heiblum,
  Mahalu, Kretinin, and Shtrikman}}]{Das2012}
\bibinfo{author}{\bibfnamefont{A.}~\bibnamefont{Das}},
  \bibinfo{author}{\bibfnamefont{Y.}~\bibnamefont{Ronen}},
  \bibinfo{author}{\bibfnamefont{M.}~\bibnamefont{Heiblum}},
  \bibinfo{author}{\bibfnamefont{D.}~\bibnamefont{Mahalu}},
  \bibinfo{author}{\bibfnamefont{A.~V.} \bibnamefont{Kretinin}},
  \bibnamefont{and}
  \bibinfo{author}{\bibfnamefont{H.}~\bibnamefont{Shtrikman}},
  \bibinfo{journal}{Nat. Comm.} \textbf{\bibinfo{volume}{3}},
  \bibinfo{pages}{1165} (\bibinfo{year}{2012}).

\bibitem[{\citenamefont{Tiwari et~al.}(2012)\citenamefont{Tiwari, Belzig, Sigrist, and Bruder}}]{Tiwari2014}
\bibinfo{author}{\bibfnamefont{R.~P.}~\bibnamefont{Tiwari}},
  \bibinfo{author}{\bibfnamefont{W.}~\bibnamefont{Belzig}},
  \bibinfo{author}{\bibfnamefont{M.}~\bibnamefont{Sigrist}},
  \bibnamefont{and}
  \bibinfo{author}{\bibfnamefont{C.}~\bibnamefont{Bruder}},
  \bibinfo{journal}{Phys. Rev. B} \textbf{\bibinfo{volume}{89}},
  \bibinfo{pages}{184512} (\bibinfo{year}{2014}).  

\bibitem[{\citenamefont{Walter et~al.}(2012)\citenamefont{Walter, Budich, Elsert, and Trauzettel}}]{Walter2013}
\bibinfo{author}{\bibfnamefont{S.}~\bibnamefont{Walter}},
  \bibinfo{author}{\bibfnamefont{J.~C.}~\bibnamefont{Budich}},
  \bibinfo{author}{\bibfnamefont{J.}~\bibnamefont{Elsert}},
  \bibnamefont{and}
  \bibinfo{author}{\bibfnamefont{B.}~\bibnamefont{Trauzettel}},
  \bibinfo{journal}{Phys. Rev. B} \textbf{\bibinfo{volume}{88}},
  \bibinfo{pages}{035441} (\bibinfo{year}{2013}).    
  
\bibitem[{\citenamefont{Cottet et~al.}(2012)\citenamefont{Cottet, Kontos, and
  Levy~Yeyati}}]{cottet12}
\bibinfo{author}{\bibfnamefont{A.}~\bibnamefont{Cottet}},
  \bibinfo{author}{\bibfnamefont{T.}~\bibnamefont{Kontos}}, \bibnamefont{and}
  \bibinfo{author}{\bibfnamefont{A.}~\bibnamefont{Levy~Yeyati}},
  \bibinfo{journal}{Phys. Rev. Lett.} \textbf{\bibinfo{volume}{108}},
  \bibinfo{pages}{166803} (\bibinfo{year}{2012}).

\bibitem[{\citenamefont{Kawabata}(2001)}]{Kawabata-2001a}
\bibinfo{author}{\bibfnamefont{S.}~\bibnamefont{Kawabata}},
  \bibinfo{journal}{J. Phys. Soc. Jpn.} \textbf{\bibinfo{volume}{70}},
  \bibinfo{pages}{1210} (\bibinfo{year}{2001}).

\bibitem[{\citenamefont{Braunecker et~al.}(2013)\citenamefont{Braunecker,
  Burset, and Levy~Yeyati}}]{Braunecker-2013a}
\bibinfo{author}{\bibfnamefont{B.}~\bibnamefont{Braunecker}},
  \bibinfo{author}{\bibfnamefont{P.}~\bibnamefont{Burset}}, \bibnamefont{and}
  \bibinfo{author}{\bibfnamefont{A.}~\bibnamefont{Levy~Yeyati}},
  \bibinfo{journal}{Phys. Rev. Lett.} \textbf{\bibinfo{volume}{111}},
  \bibinfo{pages}{136806} (\bibinfo{year}{2013}).

\bibitem[{\citenamefont{Cottet}(2014)}]{cottet14}
\bibinfo{author}{\bibfnamefont{A.}~\bibnamefont{Cottet}}
  (\bibinfo{year}{2014}), \bibinfo{note}{arXiv:1406.4666 [cond-mat.mes-hall]}.

\bibitem[{\citenamefont{Chtchelkatchev
  et~al.}(2002)\citenamefont{Chtchelkatchev, Blatter, Lesovik, and
  Martin}}]{Chtchelkatchev-2002a}
\bibinfo{author}{\bibfnamefont{N.~M.} \bibnamefont{Chtchelkatchev}},
  \bibinfo{author}{\bibfnamefont{G.}~\bibnamefont{Blatter}},
  \bibinfo{author}{\bibfnamefont{G.~B.} \bibnamefont{Lesovik}},
  \bibnamefont{and} \bibinfo{author}{\bibfnamefont{T.}~\bibnamefont{Martin}},
  \bibinfo{journal}{Phys. Rev. B} \textbf{\bibinfo{volume}{66}},
  \bibinfo{pages}{161320} (\bibinfo{year}{2002}).

\bibitem[{\citenamefont{Scher\"ubl et~al.}(2014)\citenamefont{Scher\"ubl,
  P\'alyi, and Csonka}}]{Scherubl-2014a}
\bibinfo{author}{\bibfnamefont{Z.}~\bibnamefont{Scher\"ubl}},
  \bibinfo{author}{\bibfnamefont{A.}~\bibnamefont{P\'alyi}}, \bibnamefont{and}
  \bibinfo{author}{\bibfnamefont{S.}~\bibnamefont{Csonka}},
  \bibinfo{journal}{Phys. Rev. B} \textbf{\bibinfo{volume}{89}},
  \bibinfo{pages}{205439} (\bibinfo{year}{2014}).

\bibitem[{\citenamefont{Burkard et~al.}(2000)\citenamefont{Burkard, Loss, and
  Sukhorukov}}]{Burkard-2000a}
\bibinfo{author}{\bibfnamefont{G.}~\bibnamefont{Burkard}},
  \bibinfo{author}{\bibfnamefont{D.}~\bibnamefont{Loss}}, \bibnamefont{and}
  \bibinfo{author}{\bibfnamefont{E.~V.} \bibnamefont{Sukhorukov}},
  \bibinfo{journal}{Phys. Rev. B} \textbf{\bibinfo{volume}{61}},
  \bibinfo{pages}{R16303} (\bibinfo{year}{2000}).

\bibitem[{\citenamefont{Veronica Cerletti, Oliver Gywat, and Daniel Loss}(2005)\citenamefont{Cerletti, Gywat, and
  Loss}}]{Cerletti-2005a}
\bibinfo{author}{\bibfnamefont{V.}~\bibnamefont{Cerletti}},
  \bibinfo{author}{\bibfnamefont{O.}~\bibnamefont{Gywat}}, \bibnamefont{and}
  \bibinfo{author}{\bibfnamefont{D.} \bibnamefont{Loss}},
  \bibinfo{journal}{Phys. Rev. B} \textbf{\bibinfo{volume}{72}},
  \bibinfo{pages}{115316} (\bibinfo{year}{2005}).

\bibitem[{\citenamefont{Jan Budich and Bj{\"o}rn
      Trauzettel}(2010)\citenamefont{Budich and Trauzettel}}]{Budich-2010a}
\bibinfo{author}{\bibfnamefont{J.}~\bibnamefont{Budich}}, \bibnamefont{and}
  \bibinfo{author}{\bibfnamefont{B.} \bibnamefont{Trauzettel}},
  \bibinfo{journal}{Nanotechnology} \textbf{\bibinfo{volume}{21}},
  \bibinfo{pages}{274001} (\bibinfo{year}{2010}).

\bibitem[{}]{footnote}In particular, in order to disentangle the
  electronic from the photonic degrees of freedom, the photon emitters need to be prepared in a quantum
  superposition state.

\bibitem[{\citenamefont{Brossard et all.}(2013)\citenamefont{Brossard, Reid,
  Chan, Xu, Griffiths, Williams, Murray, Taylor}}]{Brossard-2013a}
\bibinfo{author}{\bibfnamefont{F.~S.~F.}~\bibnamefont{Brossard}},
  \bibinfo{author}{\bibfnamefont{B.~P.~L.}~\bibnamefont{Reid}},
  \bibinfo{author}{\bibfnamefont{C.~C.~S.}~\bibnamefont{Chan}},
  \bibinfo{author}{\bibfnamefont{X.~L.}~\bibnamefont{Xu}},
  \bibinfo{author}{\bibfnamefont{J.~P.} \bibnamefont{Griffiths}},
  \bibinfo{author}{\bibfnamefont{D.~A.}~\bibnamefont{Williams}},
  \bibinfo{author}{\bibfnamefont{R.}~\bibnamefont{Murray}},
  \bibinfo{author}{\bibfnamefont{R.~A.} \bibnamefont{Taylor}},
  \bibinfo{journal}{Opt. Express} \textbf{\bibinfo{volume}{21}},
  \bibinfo{pages}{16934--16945} (\bibinfo{year}{2013}).

\bibitem[{\citenamefont{Schroer}(2014)}]{Schroer2014}
\bibinfo{author}{\bibfnamefont{A.}~\bibnamefont{Schroer}}, \bibnamefont{and}
\bibinfo{author}{\bibfnamefont{P.}~\bibnamefont{Recher}},
  (\bibinfo{year}{2014}), \bibinfo{note}{arXiv:1412.8619 [cond-mat.supr-con]}.
\bibitem[{\citenamefont{Kim et~al.}(2004)\citenamefont{Kim, Kim,
  Lee}}]{Kim-2006a}
\bibinfo{author}{\bibfnamefont{S.-H.}~\bibnamefont{Kim}},
  \bibinfo{author}{\bibfnamefont{S.-K.}~\bibnamefont{Kim}},
  \bibnamefont{and}
  \bibinfo{author}{\bibfnamefont{Y.-H}~\bibnamefont{Lee}},
  \bibinfo{journal}{Phys. Rev. B} \textbf{\bibinfo{volume}{73}},
  \bibinfo{pages}{235117} (\bibinfo{year}{2006}).

\bibitem[{\citenamefont{Tran et~al.}(2009)\citenamefont{Tran, Combri\'e, and
  De~Rossi}}]{Tran-2009a}
\bibinfo{author}{\bibfnamefont{N.-V.-Q.} \bibnamefont{Tran}},
  \bibinfo{author}{\bibfnamefont{S.}~\bibnamefont{Combri\'e}},
  \bibnamefont{and} \bibinfo{author}{\bibfnamefont{A.}~\bibnamefont{De~Rossi}},
  \bibinfo{journal}{Phys. Rev. B} \textbf{\bibinfo{volume}{79}},
  \bibinfo{pages}{041101} (\bibinfo{year}{2009}).

\bibitem[{\citenamefont{Portalupi et~al.}(2010)\citenamefont{Portalupi, Galli,
  Reardon, Krauss, O'Faolain, Andreani, Gerace}}]{Portalupi-2010a}
\bibinfo{author}{\bibfnamefont{S. L.}~\bibnamefont{Portalupi}},
  \bibinfo{author}{\bibfnamefont{M.}~\bibnamefont{Galli}},
  \bibinfo{author}{\bibfnamefont{C.}~\bibnamefont{Reardon}},
  \bibinfo{author}{\bibfnamefont{T. F.}~\bibnamefont{Krauss}},
  \bibinfo{author}{\bibfnamefont{L.}~\bibnamefont{O'Faolain}},
  \bibinfo{author}{\bibfnamefont{L. C.}~\bibnamefont{Andreani}},
  \bibnamefont{and}
  \bibinfo{author}{\bibfnamefont{D.}~\bibnamefont{Gerace}},
  \bibinfo{journal}{Optics Express} \textbf{\bibinfo{volume}{18}},
  \bibinfo{pages}{16064} (\bibinfo{year}{2010}).

\bibitem[{\citenamefont{Yoshie et~al.}(2004)\citenamefont{Yoshie, Scherer,
  Hendrickson, Khitrova, Gibbs, Rupper, Ell, Shchekin, and
  Deppe}}]{Yoshie-2004a}
\bibinfo{author}{\bibfnamefont{T.}~\bibnamefont{Yoshie}},
  \bibinfo{author}{\bibfnamefont{A.}~\bibnamefont{Scherer}},
  \bibinfo{author}{\bibfnamefont{J.}~\bibnamefont{Hendrickson}},
  \bibinfo{author}{\bibfnamefont{G.}~\bibnamefont{Khitrova}},
  \bibinfo{author}{\bibfnamefont{H.~M.} \bibnamefont{Gibbs}},
  \bibinfo{author}{\bibfnamefont{G.}~\bibnamefont{Rupper}},
  \bibinfo{author}{\bibfnamefont{C.}~\bibnamefont{Ell}},
  \bibinfo{author}{\bibfnamefont{O.~B.} \bibnamefont{Shchekin}},
  \bibnamefont{and} \bibinfo{author}{\bibfnamefont{D.~G.} \bibnamefont{Deppe}},
  \bibinfo{journal}{Nature} \textbf{\bibinfo{volume}{432}},
  \bibinfo{pages}{200} (\bibinfo{year}{2004}).



\bibitem[{\citenamefont{Majumdar et~al.}(2013)\citenamefont{Majumdar, Kaer,
  Bajcsy, Kim, Lagoudakis, Rundquist, and Vu\ifmmode \check{c}\else
  \v{c}\fi{}kovi\ifmmode~\acute{c}\else \'{c}\fi{}}}]{Majumdar-2013a}
\bibinfo{author}{\bibfnamefont{A.}~\bibnamefont{Majumdar}},
  \bibinfo{author}{\bibfnamefont{P.}~\bibnamefont{Kaer}},
  \bibinfo{author}{\bibfnamefont{M.}~\bibnamefont{Bajcsy}},
  \bibinfo{author}{\bibfnamefont{E.~D.} \bibnamefont{Kim}},
  \bibinfo{author}{\bibfnamefont{K.~G.} \bibnamefont{Lagoudakis}},
  \bibinfo{author}{\bibfnamefont{A.}~\bibnamefont{Rundquist}},
  \bibnamefont{and} \bibinfo{author}{\bibfnamefont{J.}~\bibnamefont{Vu\ifmmode
  \check{c}\else \v{c}\fi{}kovi\ifmmode~\acute{c}\else \'{c}\fi{}}},
  \bibinfo{journal}{Phys. Rev. Lett.} \textbf{\bibinfo{volume}{111}},
  \bibinfo{pages}{027402} (\bibinfo{year}{2013}).

\bibitem[{\citenamefont{Sweeney et~al.}(2014)\citenamefont{Sweeney, Carter,
  Bracker, Kim, Kim, Yang, Vora, Brereton, Cleveland, and
  Gammon}}]{Sweeney-2014a}
\bibinfo{author}{\bibfnamefont{T.~M.} \bibnamefont{Sweeney}},
  \bibinfo{author}{\bibfnamefont{S.~G.} \bibnamefont{Carter}},
  \bibinfo{author}{\bibfnamefont{A.~S.} \bibnamefont{Bracker}},
  \bibinfo{author}{\bibfnamefont{M.}~\bibnamefont{Kim}},
  \bibinfo{author}{\bibfnamefont{C.~S.} \bibnamefont{Kim}},
  \bibinfo{author}{\bibfnamefont{L.}~\bibnamefont{Yang}},
  \bibinfo{author}{\bibfnamefont{P.~M.} \bibnamefont{Vora}},
  \bibinfo{author}{\bibfnamefont{P.~G.} \bibnamefont{Brereton}},
  \bibinfo{author}{\bibfnamefont{E.~R.} \bibnamefont{Cleveland}},
  \bibnamefont{and} \bibinfo{author}{\bibfnamefont{D.}~\bibnamefont{Gammon}},
  \bibinfo{journal}{Nat. Photonics} \textbf{\bibinfo{volume}{8}},
  \bibinfo{pages}{442} (\bibinfo{year}{2014}).

\bibitem[{\citenamefont{Gywat et~al.}(2010)\citenamefont{Gywat, Krenner, and
  Berezovsky}}]{Gywat-2010}
\bibinfo{author}{\bibfnamefont{O.}~\bibnamefont{Gywat}},
  \bibinfo{author}{\bibfnamefont{H.}~\bibnamefont{Krenner}}, \bibnamefont{and}
  \bibinfo{author}{\bibfnamefont{J.}~\bibnamefont{Berezovsky}},
  \emph{\bibinfo{title}{Spins in Optically Active Quantum Dots}}
  (\bibinfo{publisher}{Wiley-VCH}, \bibinfo{year}{2010}).

\bibitem[{\citenamefont{Clauser et~al.}(1969)\citenamefont{Clauser, Horne,
  Shimony, and Holt}}]{Clauser-1969a}
\bibinfo{author}{\bibfnamefont{J.~F.} \bibnamefont{Clauser}},
  \bibinfo{author}{\bibfnamefont{M.~A.} \bibnamefont{Horne}},
  \bibinfo{author}{\bibfnamefont{A.}~\bibnamefont{Shimony}}, \bibnamefont{and}
  \bibinfo{author}{\bibfnamefont{R.~A.} \bibnamefont{Holt}},
  \bibinfo{journal}{Phys. Rev. Lett.} \textbf{\bibinfo{volume}{23}},
  \bibinfo{pages}{880} (\bibinfo{year}{1969}).

\bibitem[{\citenamefont{Weisskopf and Wigner}(1930)}]{Weisskopf-1930a}
\bibinfo{author}{\bibfnamefont{V.}~\bibnamefont{Weisskopf}} \bibnamefont{and}
  \bibinfo{author}{\bibfnamefont{E.}~\bibnamefont{Wigner}},
  \bibinfo{journal}{Z. Phys.} \textbf{\bibinfo{volume}{63}},
  \bibinfo{pages}{54} (\bibinfo{year}{1930}).

\bibitem[{\citenamefont{Jaynes and Cummings}(1963)}]{Jaynes-1963a}
\bibinfo{author}{\bibfnamefont{E.}~\bibnamefont{Jaynes}} \bibnamefont{and}
  \bibinfo{author}{\bibfnamefont{F.}~\bibnamefont{Cummings}},
  \bibinfo{journal}{Proc. IEEE} \textbf{\bibinfo{volume}{51}},
  \bibinfo{pages}{89 } (\bibinfo{year}{1963}).

\bibitem[{\citenamefont{Vidal and Werner}(2002)}]{Vidal-2002a}
\bibinfo{author}{\bibfnamefont{G.}~\bibnamefont{Vidal}} \bibnamefont{and}
  \bibinfo{author}{\bibfnamefont{R.~F.} \bibnamefont{Werner}},
  \bibinfo{journal}{Phys. Rev. A} \textbf{\bibinfo{volume}{65}},
  \bibinfo{pages}{032314} (\bibinfo{year}{2002}).

\bibitem[{\citenamefont{Kloeffel and Loss}(2013)}]{Kloeffel-2013}
\bibinfo{author}{\bibfnamefont{C.}~\bibnamefont{Kloeffel}} \bibnamefont{and}
  \bibinfo{author}{\bibfnamefont{D.}~\bibnamefont{Loss}},
  \bibinfo{journal}{Annu. Rev. Condens. Matter Phys.}
  \textbf{\bibinfo{volume}{4}}, \bibinfo{pages}{51} (\bibinfo{year}{2013}).

\end{thebibliography}

\end{document}